\documentstyle[12pt,epsf]{article}

\advance \topmargin by -\headheight
\advance \topmargin by -\headsep

\evensidemargin \oddsidemargin
  \def\tr{{\hbox{\rm Tr}}}

\def\ie{{\em i.e.}}
\def\ie{\hbox{\it i.e.}}

\def\CC{{\mathchoice
{\rm C\mkern-8mu\vrule height1.45ex depth-.05ex
width.05em\mkern9mu\kern-.05em}
{\rm C\mkern-8mu\vrule height1.45ex depth-.05ex
width.05em\mkern9mu\kern-.05em}
{\rm C\mkern-8mu\vrule height1ex depth-.07ex
width.035em\mkern9mu\kern-.035em}
{\rm C\mkern-8mu\vrule height.65ex depth-.1ex
width.025em\mkern8mu\kern-.025em}}}

\def\RR{{\rm I\kern-1.6pt {\rm R}}}

\def\ZZ{{\rm Z}\kern-3.8pt {\rm Z} \kern2pt}
\def\IB{\relax{\rm I\kern-.18em B}}
\def\ID{\relax{\rm I\kern-.18em D}}
\def\II{\relax{\rm I\kern-.18em I}}
\def\IP{\relax{\rm I\kern-.18em P}}

\def\np{Nucl. Phys.}
\def\pl{Phys. Lett.}
\def\prl{Phys. Rev. Lett.}
\def\pr{Phys. Rev.}

\def\rmp{Rev. Mod. Phys.}

\def\atmp{Adv. Theor. Math. Phys. }
\def\jhep{J. High Energy Phys.}
\def\ptp{Prog. Theor. Phys.}

\def\atmp{Adv. Theor. Math. Phys.}

\newcommand{\beq}{\begin{equation}}
\newcommand{\eeq}{\end{equation}}
\newcommand{\rc}{\nonumber\\}
\newcommand{\bear}{\begin{eqnarray}}
\newcommand{\eear}{\end{eqnarray}}

\def\to{\rightarrow}

\def\tr{{\rm Tr}}
\def\to{\rightarrow}

\newfont{\namefont}{cmr10}
\newfont{\addfont}{cmti7 scaled 1440}
\newfont{\boldmathfont}{cmbx10}
\newfont{\headfontb}{cmbx10 scaled 1728}
\renewcommand{\theequation}{{\rm\thesection.\arabic{equation}}}
\begin{document}
\begin{titlepage}

\begin{center} \Large \bf Adding flavor to the gravity dual of non-commutative
gauge theories

\end{center}

\vskip 0.3truein
\begin{center}
Daniel Are\'an${}^{\,*}$
\footnote{arean@fpaxp1.usc.es},
Angel Paredes${}^{\,\dagger}$
\footnote{Angel.Paredes@cpht.polytechnique.fr}
and
Alfonso V. Ramallo${}^{\,*}$
\footnote{alfonso@fpaxp1.usc.es}

\vspace{0.3in}

${}^{\,*}$Departamento de F\'\i sica de Part\'\i culas, Universidade de
Santiago de Compostela \\and\\
Instituto Galego de F\'\i sica de Altas Enerx\'\i as (IGFAE)\\
E-15782 Santiago de Compostela, Spain
\vspace{0.3in}

${}^{\,\dagger}$ Centre de Physique Th\'eorique, \'Ecole Polytechnique, 91128
Palaiseau, France

\end{center}
\vskip.5
truein

\begin{center}
\bf ABSTRACT
\end{center}
We study the addition of flavor degrees of freedom to the 
supergravity dual of the
non-commutative  deformation of the maximally supersymmetric gauge 
theories. By considering D7
flavor branes in the probe approximation and studying their 
fluctuations we extract the spectrum
of scalar and vector mesons as a function of the non-commutativity. 
We find that the spectrum
for very large non-commutative parameter is equal to the one in the 
commutative theory, while
for some intermediate values of the non-commutativity some of the 
modes disappear from the
discrete spectrum. We also study the semiclassical dynamics of 
rotating open  strings attached
to the D7-brane, which correspond to mesons with large spin. Under 
the effect of the
non-commutativity the open strings get tilted. However, at 
small(large) distances they display
the same Regge-like (Coulombic) behaviour as in the commutative 
theory. We also consider the
addition of D5-flavor branes to the non-commutative deformation of 
the ${\cal N}=1$
supersymmetric Maldacena-N\'u\~nez background.

\vskip3.6truecm
\leftline{US-FT-1/05}
\leftline{CPHT - RR 027.0505}
\leftline{hep-th/0505181 \hfill May 2005}
\smallskip
\end{titlepage}
\setcounter{footnote}{0}



\setcounter{equation}{0}
\section{Introduction}
\medskip
The gauge/gravity correspondence \cite{jm} is a consequence of the 
two complementary
descriptions of D-branes: an open string description as hyperplanes 
to which open strings are
attached and a closed string description as solitons of the type II 
low energy string effective
action \cite{MAGOO}. In the open string picture the D-brane dynamics 
is determined by (super)
Yang-Mills theory in flat space whereas, from the   closed string 
viewpoint, the D-branes are
associated to non-trivial solutions of the classical equations of 
supergravity.  By using this
correspondence one can obtain the quantum dynamics of gauge theories 
by analyzing the properties
of the corresponding solution of  classical supergravity.

In order to extend the gauge/gravity correspondence to more realistic 
scenarios we should be
able to describe theories with  matter in the fundamental 
representation. In a string theory
setup this corresponds to adding open strings to the supergravity 
description of gauge theories.
A simple way to achieve this is by considering a static quark source, 
as was done in ref.
\cite{Wilson} to compute the expectation value of the Wilson loop 
operator. Alternatively, it
was proposed in refs. \cite{KK, KKW} that one can add dynamical 
quarks to the gauge/gravity
correspondence by considering flavor branes and looking at their 
fluctuations. These flavor
branes must fill the spacetime directions of the gauge theory and are 
extended along the
holographic direction.

For the $AdS_5\times S^5$ geometry the appropriate flavor branes are 
D7-branes \cite{D3D7}. In
this setup one starts with a stack of parallel D3-branes and adds 
another stack of D7-branes
which fill the directions parallel to the worldvolume of the 
D3-branes and are separated from
them by a certain distance along the directions orthogonal to both 
sets of branes. The strings
stretched between the D3- and the D7-branes correspond to fundamental 
hypermultiplets in the
four-dimensional gauge theory. These hypermultiplets are massive, 
with the mass being
proportional to the separation of the two stacks of branes. In the 
decoupling limit the
D3-branes may be replaced by the
$AdS_5\times S^5$ geometry. Moreover, if the number of D7-branes is 
much smaller than the
number of D3-branes, we can neglect the backreaction of the flavor 
branes on the
$AdS_5\times S^5$ geometry, and treat the D7-branes as probes. The 
fluctuation modes of the
probe are then identified with the mesons, \ie\ with the low energy
excitations of the dual field theory. This program has been  carried out in
ref.
\cite{KMMW}, where the mass spectrum of the model has been obtained
analytically.  For similar analysis for
several backgrounds and probes, see refs. \cite{Sonnen}-\cite{Sakai}.

In this paper we shall address the problem of adding flavor to a 
gauge theory living on a
spacetime which has two spatial coordinates that are non-commutative. 
Such non-local theories
have a long story (see ref. \cite{NCFT} for a review) and have been 
the object of intense study
in recent years after the discovery  that they can be obtained from limits of
string theory and M-theory. Indeed, by considering string theory in 
the presence of a
Neveu-Schwarz $B$-field, and performing a suitable limit one ends up 
with a gauge theory on a
non-commutative space \cite{NCString, SW}.

In the context of the gauge/gravity correspondence it is quite 
natural to have a supergravity
solution dual to  the non-commutative gauge theory for large $N$ and 
strong 't Hooft coupling.
Actually, as pointed out in ref. \cite{MR}, this background can be 
obtained from the decoupling
limit of the type IIB supergravity solution \cite{Bound} representing 
a stack of non-threshold
bound states of  D3-branes and D1-branes (for a similar analysis in more general brane
setups see ref. \cite{AOSJ}). The corresponding 
ten-dimensional metric breaks
four-dimensional Lorentz invariance since it distinguishes between 
the coordinates of the
non-commutative plane and the other two Minkowski coordinates.
As expected, this solution has a non-vanishing Neveu-Schwarz
$B$-field directed along the non-commutative directions, as well as 
two Ramond-Ramond potentials
and a running dilaton.  This background contains  a parameter 
$\Theta$ which corresponds to the
non-commutative deformation of the corresponding gauge theory. For 
any non-vanishing value of
$\Theta$ the solution preserves sixteen supersymmetries, while for 
$\Theta=0$ it reduces to the
maximally supersymmetric $AdS_5\times S^5$ background. 

To add flavor to the (D1,D3) background we will consider a D7-brane 
probe. First of all we will
make use of  kappa symmetry to determine the static configuration of 
the probe that preserves
supersymmetry. It turns out that this embedding is the same as the 
one considered in ref.
\cite{KMMW}, but now the metric induced in the worldvolume is 
different and the brane captures a
non-vanishing Neveu-Schwarz $B$-field in its worldvolume. Next, we 
will study the fluctuations
of the scalar and worldvolume gauge field around the static 
configuration. Some of the scalar
fluctuation modes are coupled to the worldvolume electric field. The 
corresponding equations of
motion are coupled  and break Lorentz invariance. However, we will be 
able to find a set of
decoupled differential equations, which can be analyzed by means of 
several techniques.

One way to extract information from the decoupled equations of the 
fluctuations is by
transforming them into a Schr\"odinger equation. This can be done  by 
performing a simple change
of variables and the result is a wave equation for a quantum 
mechanical problem of a
one-dimensional particle moving under the action of a potential, 
whose form can be determined.
The analysis of this potential can be used to obtain the qualitative 
behaviour of the spectrum
of the fluctuation modes. It turns out that, in order to have a 
discrete spectrum for some of
the modes, the momentum along some directions must be bounded from 
above, the bound being a
function of the non-commutative parameter $\Theta$. Actually, for any 
non-vanishing value of
$\Theta$, the tower of energy levels of some modes is cut off from 
above and, in some window of
values  $\Theta_1<\Theta<\Theta_2$ of the non-commutative parameter, 
this upper bound is so
low that these modes disappear from the discrete spectrum. We have 
determined the value of
$\Theta_{1,2}$ both by using the WKB approximation and by solving 
numerically the differential
equation of the fluctuations by means of the shooting technique. 
Moreover, if $\Theta$ is large
enough the spectrum of fluctuations reduces to the one corresponding 
to the commutative theory,
a fact that is reminiscent of the Morita equivalence of algebras in 
the non-commutative torus
(see \cite{NCFT} for a review and further references).

In order to have a complementary picture of the meson spectrum we 
have studied the classical
dynamics of a rotating string whose ends are attached to the D7 
flavor brane. The profile of the
string can be obtained by solving the Nambu-Goto equations of motion 
in the background under
consideration with appropriate boundary conditions. We will verify 
that, as compared to the
string in the $AdS_5\times S^5$ geometry, the non-commutative 
deformation results in a tilting
of the string. The energy spectrum has a Regge-like behaviour for low 
angular momentum, with an
asymptotic Regge slope which, remarkably,  is the same as in the 
commutative case. For large
angular momentum the energy obtained is that corresponding to two 
non-relativistic masses bound
by a Coulomb potential. We will confirm these behaviours by an 
explicit calculation of the static
potential energy from a hanging string attached to the flavor brane.
As an aside, we also consider a moving hanging string and briefly comment
on some effects related to the breaking of Lorentz symmetry.

We will also explore the effect of the non-commutativity on other 
background, namely the
Maldacena-N\'u\~nez solution \cite{MN, CV}. The non-commutative 
version of this background was
found in ref. \cite{NCMN}. This solution, which corresponds to a 
(D3,D5) bound state  with the
D3- and D5-branes wrapped on a two-cycle, preserves four 
supersymmetries and is dual to
non-commutative ${\cal N}=1$ super Yang-Mills in four dimensions. One 
of the interesting
features of this solution is that the dilaton reaches a finite value 
at the UV boundary and,
therefore, one could try to use the non-commutative deformation  to 
solve some of the
difficulties of the Maldacena-N\'u\~nez background due to the blowing 
up of the dilaton at the
UV boundary. The flavor branes for the  Maldacena-N\'u\~nez geometry 
are D5-branes wrapped on a
calibrated two-dimensional submanifold \cite{WangHu}, which preserve 
the same supersymmetries as
the background. These calibrated submanifolds were explored 
systematically in ref.
\cite{flavoring} by using kappa symmetry. In this paper we will determine them
for the non-commutative version of the model. After obtaining the 
Killing spinors
of the background, we will find that the supersymmetric embeddings of 
the D5-brane probe are
exactly the same as the ones corresponding to the commutative 
background. Next, we will study
the fluctuations of the D5-brane probe around the static embeddings 
and we will discover that
there also exists a coupling between the scalar and vector 
fluctuations, as happened in the case
of a D7-brane probe in the (D1,D3) background. As expected, this 
coupling breaks Lorentz
invariance. However, the differential equations for the
coupled fluctuations are in this case much more complicated and we 
will not try to integrate
them.

This paper is organized as follows. In section 2 we will review the 
main features of the
supergravity dual of non-commutative gauge theories. In section 3 we 
will start to study the
dynamics of a D7-brane probe in this background. First we will 
determine the static
supersymmetric embedding of the probe and then we will study  its 
fluctuations. The
corresponding meson spectrum will be obtained in section 4. In 
section 5 we will perform a
semiclassical analysis of a rotating string attached to the flavor 
brane in the non-commutative
background. The calculation of  the 
corresponding static potential energy is done
in section 6.
 In section 7 we summarize our results and draw some 
conclusions. The paper contains
two appendices. In appendix A we apply the WKB method to obtain the 
energy levels of the meson
fluctuations. In appendix B we add a flavor brane to the 
non-commutative Maldacena-N\'u\~nez
background.

\setcounter{equation}{0}
\section{The supergravity dual of non-commutative gauge theories}
\medskip
In this section we will review the main features of the gravity dual 
of non-commutative gauge
theories obtained in ref.  \cite{MR} (see also ref. \cite{Bound}). 
This supergravity background
is obtained by taking the decoupling limit of the solution 
corresponding to the non-threshold
bound state of a D3-brane and a D1-brane and contains a $B$ field 
along two of the spatial
worldvolume directions of the D3-brane.  The string frame metric 
takes the form:
\beq
ds^2\,=\,{r^2\over R^2} \,[\,dx_{0,1}^2\,+\,h
dx^2_{2,3}\,]\,+\,{R^2\over r^{2}}\, [\,(dy^1)^2\,+\cdots\,+(dy^6)^2\,]\,\,,
\label{MRmetric}
\eeq
where $R^4=4\pi g_s N(\alpha')^2$ with $N$ being the number of 
D3-branes. The radial coordinate
$r$ is given by  $r^2=(y^1)^2\,+\cdots+\,(y^6)^2$ and the function $h$ is:
\beq
h\,=\,{1\over 1+\Theta^4r^4}\,\,,
\eeq
with $\Theta$ being a constant.  The dilaton is
\beq
e^{2\phi}\,=\,h\,g_s^2\,\,,
\label{MRdilaton}
\eeq
where $g_s$ is the string coupling constant.  Notice that the 
function $h$ distinguishes in the
metric the non-commutative plane $x^2x^3$ from the two other 
Minkowski directions. Moreover,
this same function is responsible of the running of the dilaton in 
eq. (\ref{MRdilaton}).
Obviously, when $\Theta=0$ the dilaton is constant and  we recover 
the $AdS_5\times S^5$ metric.
When $\Theta\not=0$ this background is dual to a gauge theory in 
which the coordinates $x^2$ and
$x^3$ do not commute, being $[x^2, x^3]\sim\Theta^2$.

This supergravity solution
also contains a NSNS three-form $H$, a RR three-form $F^{(3)}$ and a 
self-dual RR
five-form $F^{(5)}$. The expressions of the three-forms $H$ and $F^{(3)}$ are:
\bear
&&H\,=\,-{\Theta^2\over R^2}\,\partial_{y^i}\,[r^4h]\,dx^2\wedge dx^3\wedge
dy^i\,\,,\rc\rc &&g_s\,F^{(3)}\,=\,4\Theta^2\,{r^2\over R^2}\,\, y^i\,
dx^0\wedge dx^1\wedge dy^i\,\,,
\eear
while the RR five-form $F^{(5)}$ can be represented as
\beq
F^{(5)}\,=\,\hat F^{(5)}\,+\,{}^*\hat F^{(5)}\,\,,
\eeq
with $\hat F^{(5)}$ being given by:
\beq
g_s\,\hat F^{(5)}\,=\,4\,{r^2\over R^4}\,h\,y^i\,dx^0\wedge\cdots \wedge
dx^3\wedge dy^i\,\,.
\eeq
These forms can be represented in terms of
the NSNS $B$ field and the RR potentials $C_{(2)}$ and $C_{(4)}$ as follows:
\beq
H\,=\,dB\,\,,
\,\,\,\,\,\,\,\,\,\,\,\,\,\,\,\,
F^{(3)}\,=\,dC^{(2)}\,\,,
\,\,\,\,\,\,\,\,\,\,\,\,\,\,\,\,
F^{(5)}\,=\,dC^{(4)}\,-\,H\wedge C^{(2)}\,\,.
\eeq
The explicit form of these potentials is:
\bear
&&B\,=\,-\Theta^2\,{r^4\over R^2}\,h\,dx^2\wedge dx^3\,\,,\rc\rc
&&g_s\,C^{(2)}\,=\,a^2\,{r^4\over R^2} \,dx^0\wedge dx^1\,\,,\rc\rc
&&C^{(4)}\,=\hat C^{(4)}+\,\tilde C^{(4)}\,\,,
\label{potentials}
\eear
where
\beq
g_s\,\hat C^{(4)}\,=\,h\,{r^4\over R^4}\,\,dx^0\wedge\cdots \wedge dx^3\,\,,
\label{C4}
\eeq
and $\tilde C^{(4)}$ is a potential for ${}^*\hat F^{(5)}$, \ie\ ,
${}^*\hat F^{(5)}=d\tilde C^{(4)}$. From now on we will take, for 
simplicity, $g_s\,=1$.
Notice that, as expected, the $B$-field has only components
along the non-commutative plane $x^2x^3$, which approach a constant 
as $r$ becomes large (\ie\
for large energies in the gauge theory). However, for large $r$ the 
metric changes drastically
with respect to the $AdS_5\times S^5$ geometry, since the  $x^2x^3$ 
directions collapse in this
limit.

The background described above preserves sixteen supersymmetries. The 
corresponding
Killing spinors have been obtained in ref. \cite{GG}. In the natural
frame for the coordinate system we are using, namely:
\beq
e^{x^{0,1}}\,=\,{r\over R}\,dx^{0,1}\,\,,
\quad\quad\quad\quad
e^{x^{2,3}}\,=\,{r\over R}\,h^{{1\over 2}}\,dx^{2,3}\,\,,
\quad\quad\quad\quad
e^{y^{i}}\,=\,{R\over r}\,dy^{i}\,\,,
\eeq
they can  be represented as
\beq
\epsilon\,=\,e^{-{\beta\over 2}\Gamma_{x^2x^3}\sigma_3}\,\,\tilde\epsilon\,\,,
\eeq
where $\beta$ is an angle determined by the equation \cite{GG}
\beq
\cos\beta\,=\,h^{{1\over 2}}\,\,\,,
\,\,\,\,\,\,\,\,\,\,\,\,\,\,\,\,\,\,\,\,\,\,\,\,\,\,\,\,\,\,\,\,
\sin\beta\,=\,\Theta^2\,r^2\,h^{{1\over 2}}\,\,,
\label{MRbeta}
\eeq
and $\tilde\epsilon$ is the spinor satisfying the projection
\beq
\Gamma_{x^0\cdots x^3}\,(i\sigma_2)\,\tilde\epsilon\,=\,-\tilde\epsilon\,\,.
\label{tildeprojection}
\eeq
 From the above equations it is immediate to prove that
\beq
\Gamma_{x^0\cdots x^3}\,(i\sigma_2)\,\epsilon\,=\,-\,
e^{\beta\,\Gamma_{x^2x^3}\sigma_3}\,\,\epsilon\,\,.
\label{MRprojection}
\eeq

\setcounter{equation}{0}
\section{D7-brane probe}
Let us consider a D7-brane probe in the background described in 
previous section and let $\xi^m$
($m=0,\cdots 7$) be a set of worldvolume coordinates. If $X^M$ denote 
ten-dimensional
coordinates, the D7-brane embedding will be determined by the 
functions $X^M(\xi^m)$. The
induced metric in the worldvolume is
\beq
g_{mn}\,=\,\partial_{m}X^M\,\partial_{n}X^N\,G_{MN}\,\,,
\label{inducedg}
\eeq
where $G_{MN}$ is the ten-dimensional metric (\ref{MRmetric}). We are 
interested in
configurations in which the D7-brane fills the spacetime directions 
$x^0\cdots x^3$ of the gauge
theory. Therefore, it is quite natural to choose the following set of 
worldvolume coordinates
\beq
\xi^m\,=\,(x^0,\cdots,x^3,y^1,\cdots,y^4)\,\,,
\eeq
and consider embeddings in which the remaining two coordinates $y^5$ 
and $y^6$ depend on the
$\xi^m$'s. Notice that the norm of the vector $(y^5, y^6)$ determines 
the distance between the
D7-brane and the D3-branes of the background. We will consider first 
the case in which this norm is
constant along the worldvolume of the D7-brane and, in general, different
from zero. By looking at the  kappa symmetry condition of the
probe \cite{bbs}, we will show that these 
configurations are supersymmetric.
Next, we will study in detail the fluctuations around these 
embeddings and we will be able to
determine the corresponding meson spectrum.

\subsection{Kappa Symmetry}
The supersymmetric configurations of a D-brane probe in a given 
background are those for which
the following condition
\beq
\Gamma_{\kappa}\,\epsilon\,=\,\epsilon\,\,,
\label{kappa}
\eeq
is satisfied \cite{bbs}. In eq. (\ref{kappa}), $\Gamma_{\kappa}$ is a 
matrix whose explicit
expression depends on the embedding of the probe (see below) and 
$\epsilon$ is a Killing spinor
of the background. In order to write the form of $\Gamma_{\kappa}$, 
let us define the induced
worldvolume gamma matrices as
\beq
\gamma_{m}\,=\,\partial_{m}\,X^M\,E^{\bar N}_{M}\,\Gamma_{\bar N}\,\,,
\label{inducedgamma}
\eeq
where $\Gamma_{\bar N}$ are constant ten-dimensional Dirac matrices 
and $E^{\bar N}_{M}$ is the
vielbein for the metric $G_{MN}$. Then, if $\gamma_{m_1m_2\cdots}$ 
denotes the antisymmetrized
product of the induced gamma matrices (\ref{inducedgamma}), the kappa 
symmetry matrix for a
Dp-brane in the type IIB theory is \cite{swedes}:
\bear
\Gamma_{\kappa}\,=\,{1\over \sqrt{-\det(g+{\cal F})}}
&&\sum_{n=0}^{\infty}\,{(-1)^n\over 2^n n!}\,\gamma^{m_1n_1}\,
\cdots\gamma^{m_n n_n}\,\,
{\cal F}_{m_1 n_1}\,\cdots\,{\cal F}_{m_n n_n}\,\,
\times\rc\rc
&&\times(\sigma_3)^{{p-3\over 2}\,-\,n}\,\, (i\sigma_2)\,\,\Gamma_{(0)}\,\,,
\label{generalgammak}
\eear
where $g$ is the induced metric (\ref{inducedg}), $\Gamma_{(0)}$ denotes
\beq
\Gamma_{(0)}\,=\,{1\over (p+1)!}\,\,\epsilon^{m_1\cdots m_{p+1}}\,\,
\gamma_{m_1\cdots m_{p+1}}\,\,,
\label{Gammazero}
\eeq
and ${\cal F}$ is the following combination of the worldvolume gauge 
field strength $F$ and the
pullback $P[B]$ of the NSNS B field:
\beq
{\cal F}\,=\,F\,-\,P[B]\,\,.
\eeq
In eq. (\ref{generalgammak}) $\sigma_2$ and $\sigma_3$ are Pauli 
matrices that act on the two
Majorana-Weyl components (arranged as a two-dimensional vector) of 
the type IIB spinors.

Let us now consider  the  embeddings in which $y^5$ and $y^6$ are 
constant and the worldvolume
gauge  field $F$ is zero.  The induced metric for this configuration
will be denoted by $g^{(0)}$ and the value of ${\cal F}$ in this case is:
\beq
{\cal F}^{(0)}\,=\,-P[B]\,=\,\Theta^2\,{r^4\over 
R^2}\,\,h\,dx^2\wedge dx^3\,\,.
\label{MRinducedB}
\eeq
After a short calculation one can prove that the matrix 
$\Gamma_{(0)}$ for this embedding is:
\beq
\Gamma_{(0)}\,=\,h\,\Gamma_{x^0\cdots x^3}\,\Gamma_{y^1\,\cdots \,y^4}\,\,.
\eeq
Moreover, by using the expression of $g^{(0)}$, ${\cal F}^{(0)}$ and 
$h$, one gets
\beq
-\det \big(\,g^{(0)}\,+\,{\cal F}^{(0)}\big)\,=\,h\,\,.
\eeq
Then, substituting these results in the expression of $\Gamma_{\kappa}$ (eq.
(\ref{generalgammak})), one arrives at:
\beq
\Gamma_{\kappa}\,=\,h^{{1\over 2}}\,\Gamma_{y^1\,\cdots \,y^4}\,
\big[\,\Gamma_{x^0\cdots x^3}\,(i\sigma_2)\,+\,\Theta^2\,r^2\,\Gamma_{x^0x^1}\,
\sigma_1\,\big]\,\,,
\eeq
which can be rewritten as
\beq
\Gamma_{\kappa}\,=\,\Gamma_{y^1\,\cdots \,y^4}\,
e^{-\beta\,\Gamma_{x^2x^3}\sigma_3}\,\Gamma_{x^0\cdots x^3}\,(i\sigma_2)\,\,,
\eeq
where $\beta$ is the angle defined in eq. (\ref{MRbeta}).

 From this expression  and the equation (\ref{MRprojection}) satisfied 
by the Killing spinors, it
is clear that the condition $\Gamma_{\kappa}\epsilon=\epsilon$ is equivalent to
\beq
\Gamma_{y^1\,\cdots \,y^4}\,\epsilon\,=\,-\epsilon\,\,,
\eeq
which, taking into account that $[\Gamma_{y^1\,\cdots \,y^4}, 
\Gamma_{x^2x^3}]=0$,
can be put as
\beq
\Gamma_{y^1\,\cdots \,y^4}\,\tilde\epsilon\,=\,-\tilde\epsilon\,\,.
\eeq
Notice that this condition is compatible with the one written in eq.
(\ref{tildeprojection}). Thus, this embedding with $y^5$ and $y^6$ 
constant, vanishing
worldvolume gauge field  and induced
$B$ field   as in eq. (\ref{MRinducedB}), preserves eight 
supersymmetries and is
$1/4$-supersymmetric.

\subsection{Fluctuations}

Let us now consider fluctuations of the scalars  $\vec y\equiv 
(y^5,y^6)$ and of the worldvolume
gauge field $A_m$ around the configuration with $\vec 
y^{\,\,2}=(y^5)^2+(y^6)^2=L^2$ and $A_m=0$.
The lagrangian of the D7-brane probe contains two pieces:
\beq
{\cal L}\,=\,{\cal L}_{BI}\,+\,{\cal L}_{WZ}\,\,.
\eeq
The Born-Infeld term is:
\beq
{\cal L}_{BI}\,=\,-e^{-\phi}\,\sqrt{-\det(g+{\cal F})}\,\,,
\label{BI}
\eeq
while the Wess-Zumino part is
\beq
{\cal L}_{WZ}\,=\,{1\over 2}\,P[C^{(4)}]\wedge {\cal F} \wedge {\cal F}\,+\,
{1\over 6}\,P[C^{(2)}]\wedge {\cal F} \wedge {\cal F}\wedge {\cal F}\,\,.
\label{WZ}
\eeq
Let us analyze first the  Born-Infeld term. We shall expand the determinant
in (\ref{BI})  up to quadratic terms in the fluctuations. With this 
purpose in mind let us put
\beq
g\,=\,g^{(0)}\,+\,g^{(1)}\,\,,
\,\,\,\,\,\,\,\,\,\,\,\,\,\,\,\,\,\,
{\cal F}\,=\,{\cal F}^{(0)}\,+\,{\cal F}^{(1)}\,\,,
\eeq
where $g^{(0)}$ is the induced metric of the unperturbed configuration,
${\cal F}^{(0)}\,=\,-P[B]$ and
\beq
g^{(1)}_{mn}\,=\,{R^2\over r^2}\,\partial_{m}\vec y\cdot 
\partial_{n}\vec y\,\,,
\,\,\,\,\,\,\,\,\,\,\,\,\,\,\,\,\,\,
{\cal F}^{(1)}\,=\,F\,\,.
\eeq
The Born-Infeld  determinant can be written as
\beq
\sqrt{-\det(g+{\cal F})}\,=\,\sqrt{-\det \big(\,g^{(0)}\,+\,{\cal 
F}^{(0)}\,\big)}\,
\sqrt{\det(1+X)}\,\,,
\label{detX}
\eeq
where the matrix $X$ is given by:
\beq
X\,\equiv\,\bigg(\,g^{(0)}\,+\,{\cal F}^{(0)}\,\bigg)^{-1}\,\,
\bigg(\,g^{(1)}\,+\,{\cal F}^{(1)}\,\bigg)\,\,.
\eeq
To evaluate  the right-hand side of eq. (\ref{detX}), we shall use 
the expansion
\beq
\sqrt{\det(1+X)}\,=\,1\,+\,{1\over 2}\,\tr X\,-\,{1\over 4}\,\tr X^2\,+\,
{1\over 8}\,\big(\tr X\big)^2\,+\,o(X^3)\,\,.
\label{expansion}
\eeq
In order to write the terms resulting from this expansion in a neat 
form, let us introduce
the auxiliary metric
\beq
d\hat s^{\,2}\,\equiv\,{\cal G}_{mn}\,d\xi^m\,d\xi^n\,=\,
{\rho^2\,+\,L^2\over R^2}\,\big(-(dx^0)^2\,+\,\cdots\,+\,(dx^3)^2\big)\,+\,
{R^2\over \rho^2+L^2}\,\big((dy^1)^2\,+\,\cdots\,+\,(dy^4)^2\big)\,\,,
\eeq
with  $\rho^2$ being defined as:
\beq
\rho^2\,=\,(y^1)^2\,+\,\cdots\,+\,(y^4)^2\,=\,r^2\,-\,L^2\,\,.
\label{defrho}
\eeq
Notice that the metric ${\cal G}$ is nothing but the induced metric 
for the unperturbed
embeddings in the commutative ($\Theta=0$) geometry. Moreover, it 
turns out that the matrix
$\big(\,g^{(0)}\,+\,{\cal F}^{(0)}\,\big)^{-1}$ can be written in 
terms of the inverse
metric ${\cal G}^{-1}$ as:
\beq
\big(\,g^{(0)}\,+\,{\cal F}^{(0)}\,\big)^{-1}\,=\,{\cal G}^{-1}\,+\,
{\cal J}\,\,,
\label{openmetric}
\eeq
where ${\cal J}$ is an antisymmetric matrix whose only non-vanishing 
values are:
\beq
{\cal J}^{x^2x^3}\,=\,-{\cal J}^{x^3x^2}\,=\,-\Theta^2\,R^2\,\,.
\eeq
We are going to use this representation of $\big(\,g^{(0)}\,+\,{\cal 
F}^{(0)}\,\big)^{-1}$
to obtain the traces appearing in the expansion
(\ref{expansion}). Up to quadratic order in  the fluctuations, one gets:
\bear
&&\tr X\,=\,{\cal 
G}^{mn}\,g_{mn}^{(1)}\,+\,2\Theta^2\,R^2\,F_{x^2x^3}\,\,,\rc\rc
&&\tr X^2\,=\,F_{mn}\,F^{nm}\,+\,2\Theta^4\,R^4\,\big(F_{x^2x^3}\big)^{2}\,\,,
\eear
where $F^{mn}={\cal G}^{mp}{\cal G}^{nq}F_{pq}$. Using this result 
one can prove that:
\beq
e^{-\phi}\,\sqrt{-\det(g+{\cal F})}\,=\,1\,+\,\Theta^2\,R^2\,F_{x^2x^3}\,+\,
{1\over 2}\,{\cal G}^{mn}\,g_{mn}^{(1)}\,+\,{1\over 4}\,F_{mn}\,F^{mn}\,\,.
\eeq
Notice the quadratic terms are covariant with respect to the metric 
${\cal G}$. Dropping
the constant and linear terms (which do not contribute to the 
equations of motion), we
have:
\beq
{\cal L}_{BI}\,=\,-\sqrt{-\det {\cal G}}\,\bigg[\,{R^2\over 2(\rho^2+L^2)}\,
{\cal G}^{mn}\,\partial_{m}\vec y\cdot \partial_{n}\vec y\,+\,
{1\over 4}\,F_{mn}\,F^{mn}\,\bigg]\,\,,
\label{quadraticBI}
\eeq
where we have included the factor $\sqrt{-\det {\cal G}}$ (which is 
one in the coordinates
$(x^0,\cdots,x^3,y^1,\cdots,y^4)$). Remarkably, the Born-Infeld 
lagrangian written above is
exactly the same as that of the fluctuations in the $\Theta=0$ 
geometry and, in particular, is
Lorentz invariant in the Minkowski directions $x^0\cdots x^3$. This 
is so because there is a
conspiracy between the terms of the metric (\ref{MRmetric}) that 
break Lorentz symmetry in the
$x^0\cdots x^3$ coordinates and the
$B$ field which results in the same quadratic Born-Infeld lagrangian 
for the fluctuations as in
the commutative theory. Actually, for $\rho>>L$ the effective metric 
${\cal G}$ approaches that
of the $AdS_5\times S^3$ space, which can be interpreted as the fact 
that the conformal
invariance, broken by the mass of the hypermultiplet and the 
non-commutativity, is restored at
asymptotic energies.

It is interesting at this  point to recall from  the analysis of ref.
\cite{SW} that the metric relevant for the non-commutative gauge theory is
the so-called open string metric, which is the effective metric seen by
the open strings and should not be confused with the closed string
metric.  By comparing the commutative formalism with the $B$
field and the non-commutative description without the $B$ field one can
find the open string metric in terms of the closed string metric and the 
$B$ field. Moreover,  one can establish the so-called Seiberg-Witten
map between the non-commutative and commutative gauge fields \cite{SW}.

It has been proposed in ref. \cite{LiWu} that the metric
(\ref{MRmetric}) of the supergravity dual should be understood as the
closed string metric. We would like to argue here that  ${\cal G}$ is the
open string metric relevant for our problem. First of all,  notice that
we are identifying our fluctuations with open string degrees of freedom
and, thus, it is natural to think that the metric governing their
dynamics is not necessarily the same as that of the closed string
background. Moreover, ${\cal G}$ certainly contains the effect of the
coupling to the $B$ field and, actually (see  eq. (\ref{openmetric})), ${\cal G}^{-1}$
is the symmetric part of $(\,g^{(0)}\,-\,P[B]\,)^{-1}$, in  agreement
with the expression of the open string metric given in ref. \cite{SW}. Notice also that
we are keeping quadratic terms in our expansion, which is enough to study the mass
spectrum. Higher order terms, which represent interactions, are of course dependent on
$\Theta$. This fact is consistent with the Seiberg-Witten map since the
$\Theta$-dependent $*$-product can be replaced by the ordinary
multiplication in the quadratic terms of the action\footnote{It is also
interesting to consider the action of a D3-brane probe extended along
$x^0\cdots x^3$ in the (D1,D3) background. By expanding the Born-Infeld
action one can verify that the metric appearing in the quadratic
terms of the lagrangian for the fluctuations is also
$\Theta$-independent and Lorentz invariant in the directions $x^0\cdots
x^3$.}.

Let us consider next the Wess-Zumino term of the lagrangian, written 
in eq. (\ref{WZ}). First of
all,  since
$P[B]\wedge P[B]=P[\hat C^{(4)}]\wedge P[B]=0$, one can write ${\cal 
L}_{WZ}$, at
quadratic order in the fluctuations, as:
\beq
{\cal L}_{WZ}\,=\,{1\over 2}\,P[\,{\cal C}^{(4)}\,]\wedge F\wedge F\,+\,
{1\over 2}\,P[\,{\tilde C}^{(4)}\,]\wedge F\wedge F\,-\,
\,P[\,{\tilde C}^{(4)}\,]\wedge F\wedge B\,\,,
\label{finalWZ}
\eeq
where the four-form ${\cal C}^{(4)}$ is defined as:
\beq
{\cal C}^{(4)}\,\equiv\,\hat C^{(4)}\,-\,C^{(2)}\wedge B\,\,.
\eeq
A straightforward calculation, using the expressions of $\hat C^{(4)}$,
$C^{(2)}$ and $B$ written in eqs. (\ref{potentials}) and (\ref{C4}), 
gives the value of
${\cal C}^{(4)}$, namely:
\beq
{\cal C}^{(4)}\,=\,{r^4\over R^4}\,dx^0\wedge\cdots \wedge dx^3\,\,.
\label{calC4}
\eeq
Notice that the first term  of ${\cal L}_{WZ}$ in eq. (\ref{finalWZ}) 
and the expression
of the potential ${\cal C}^{(4)}$ displayed in eq. (\ref{calC4}), are 
exactly the same as
those corresponding to the D7-brane probes in the $\Theta=0$ geometry 
\cite{KMMW}.
In order to find out the contribution of the other two terms in eq. 
(\ref{finalWZ}),
let us determine the explicit form of ${\tilde C}^{(4)}$. With this 
purpose, let us
choose new coordinates, such that:
\beq
(dy^1)^2\,+\cdots+(dy^4)^2\,=\,d\rho^2\,+\,\rho^2\,d\Omega_3^2\,\,,
\eeq
where $d\Omega_3^2$ is the line element of a three-sphere of unit 
radius. In this system
of coordinates, one has:
\beq
{}^*\hat F^{(5)}\,=\,{4\rho^4\,R^4\over (\,\rho^2\,+\,\vec
y^{\,2}\,)^3}\,\,\omega_3\wedge  dy^5\wedge dy^6\,+\,{4\rho^3\,R^4\over
(\,\rho^2\,+\,\vec y^{\,2}\,)^3}\, d\rho\wedge \omega_3\wedge
(\,y^5dy^6\,-\,y^6dy^5\,)\,\,,
\eeq
where $\omega_3$ is the volume form of the three-sphere. It is not 
difficult now to obtain
the expression of a four-form  ${\tilde C}^{(4)}$ such that ${}^*\hat
F^{(5)}\,=\,d{\tilde C}^{(4)}$, namely:
\beq
{\tilde C}^{(4)}\,=\,-\,R^4\,{2\rho^2\,+\,\vec y^{\,2}\over (\,\rho^2\,+\,\vec
y^{\,2}\,)^2}\,
\omega_3\wedge (\,y^5dy^6\,-\,y^6dy^5\,)\,\,.
\eeq
Let us consider next, without loss of generality,  the fluctuations 
around the configuration with
$\vec y=(0, L)$ and, following ref. \cite{KMMW}, let us write:
\beq
y^5\,=\,\varphi\,\,,
\,\,\,\,\,\,\,\,\,\,\,\,\,\,\,\,\,\,
y^6\,=\,L\,+\,\chi\,\,.
\eeq
Then, at first order in the fluctuations ${\tilde C}^{(4)}$ can be written as:
\beq
{\tilde C}^{(4)}\,=\,L\,R^4\,{2\rho^2\,+\,L^{2}\over 
(\,\rho^2\,+\,L^{2}\,)^2}\,
\omega_3\wedge d\varphi\,\,.
\label{tildeC4}
\eeq
It is immediate from this expression that the second term in 
(\ref{finalWZ}) is negligible
at second order. Let us denote by $\tilde{\cal L}_{WZ}$ the third 
term in ${\cal L}_{WZ}$. By
plugging the value of ${\tilde C}^{(4)}$ given in eq. (\ref{tildeC4}) 
one easily obtains
$\tilde{\cal L}_{WZ}$, namely:
\beq
\tilde{\cal L}_{WZ}\,=\,-\Theta^2\,\,R^2\,L\,(\,2\rho^2\,+\,L^2\,)\,h(\rho)\,
\sqrt{\det \tilde g}\,\big(\,\partial_{\rho}\varphi\,F_{x^0x^1}\,+\,
\partial_{x^0}\varphi\,F_{x^1\rho}\,+\,\partial_{x^1}\varphi\,F_{\rho 
x^0}\,\big)\,\,,
\eeq
where $\tilde g$ is the metric of the $S^3$ and
$h(\rho)^{-1}\,=\,1\,+\,\Theta^4\,(\,\rho^2\,+\,L^2\,)^2$. By 
integrating by parts,
and using the Bianchi identity, $\partial_{\rho}F_{x^0x^1}\,+\,
\partial_{x^0}\,F_{x^1\rho}\,+\,\partial_{x^1}\,F_{\rho 
\,x^0}\,=\,0$, one can write a more
simplified  expression for $\tilde{\cal L}_{WZ}$, namely:
\beq
\tilde{\cal L}_{WZ}\,=\,R^2\,\, f(\rho)\,\sqrt{\det \tilde 
g}\,\,\,\varphi F_{x^0x^1}\,\,,
\label{WZfluct}
\eeq
where we have defined the function
\beq
f(\rho)\equiv
\Theta^2\,L\,\partial_{\rho}\,\Big[\,(\,2\rho^2\,+\,L^2\,)\,h(\rho)\,\Big]\,\,.
\eeq
Thus, the effect of the non-commutativity on the fluctuations is just 
the introduction of a
coupling between the scalar fluctuation $\varphi$ and the gauge field 
components $A_{x^0}$
and $A_{x^1}$. Notice that $f(\rho)\to 0$ as $\rho\to \infty$, which 
means that the scalar and
vector fluctuations decouple at the UV boundary and, thus, they 
behave as in the commutative
theory when $\rho$ is large.

The equations of motion for the other scalar fluctuation $\chi$ and for the
remaining components of the gauge field are the same as in the 
$\Theta=0$ case (see \cite{KMMW}
for a detailed analysis). Here we will concentrate on the study of 
the fluctuations that depend
on the non-commutative parameter $\Theta$ which, as follows from the 
above equations,  are
those corresponding to the fields $\varphi$, $A_{x^0}$ and $A_{x^1}$. 
As argued in ref.
\cite{KMMW}, by imposing the condition 
$\partial^{x^\mu}\,A_{x^\mu}=0$ one can consistently put
to zero the gauge field components along the three-sphere and the 
radial coordinate $\rho$.
These are the  modes called of type II in ref. \cite{KMMW}, which 
here mix with the fluctuations
along the scalar $\varphi$. The corresponding  equations of motion are:
\bear
&& {R^4\over (\rho^2+L^2)^2}\,\partial^{\mu}\partial_{\mu}\,A_{x^0}\,+\,{1\over
\rho^3}\,
\partial_{\rho}\big(\rho^3\partial_{\rho} A_{x^0}\big)\,+\,
{1\over \rho^2}\,\nabla^i\nabla_i\,A_{x^0}\,-\,R^2\,{f(\rho)\over
\rho^3}\,\partial_{x^1}\,\varphi\,=\,0\,\,,\rc\rc
&& {R^4\over (\rho^2+L^2)^2}\,\partial^{\mu}\partial_{\mu}\,A_{x^1}\,+\,{1\over
\rho^3}\,
\partial_{\rho}\big(\rho^3\partial_{\rho} A_{x^1}\big)\,+\,
{1\over \rho^2}\,\nabla^i\nabla_i\,A_{x^1}\,-\,R^2\,{f(\rho)\over
\rho^3}\,\partial_{x^0}\,\varphi\,=\,0\,\,,\rc\rc
&& {R^4\over (\rho^2+L^2)^2}\,\partial^{\mu}\partial_{\mu}\,\varphi\,+\,{1\over
\rho^3}\,
\partial_{\rho}\big(\rho^3\partial_{\rho} \varphi\big)\,+\,
{1\over \rho^2}\,\nabla^i\nabla_i\,\varphi\,+\,R^2\,{f(\rho)\over
\rho^3}\,F_{x^0x^1}\,=\,0\,\,.
\label{typeII}
\eear
Notice that, as the scalar $\varphi$ decouples from the equation of
$-\partial_{x^0}\,A_{x^0}+\partial_{x^1}\,A_{x^1}$,
the Lorentz condition $\partial^{x^\mu}\,A_{x^\mu}=0$  is consistent with the
equations written above.

\subsection{Decoupling of the equations}

The introduction of some non-commutativity parameter $[x_\nu,x_\nu]
\sim \theta_{\mu\nu}$ explicitly breaks the Lorentz group
$SO(1,3) \to SO(1,1) \times SO(2)$. Therefore, one can build two
Casimir operators related to $p_\mu p^\mu$ and $p_\mu \theta^{\mu\nu}
p_\nu$ \cite{gaume}. Accordingly, let us define the operator:
\beq
{\cal P}^2\,\equiv\,-\partial^2_{x^0}\,+\,\partial^2_{x^1}\,\,.
\eeq
Moreover, the squared mass $M^2$ is the eigenvalue of the 
operator
$\partial^{\mu}\partial_{\mu}=-\partial^2_{x^0}+\partial^2_{x^1}+\partial^2_{x^2}
+\partial^2_{x^3}\ \ $. Let us use these operators  to decouple the
equations (\ref{typeII}). First of all,  we can combine  the equations of
$A_{x^0}$ and $A_{x^1}$ to get an  equation for the field
strength $F_{x^0x^1}$. The equation for $F_{x^0x^1}$ becomes:
\beq
{R^4\over (\rho^2+L^2)^2}\,\partial^{\mu}\partial_{\mu}\,F_{x^0x^1}\,+\,
{1\over \rho^3}\partial_{\rho}\big(\rho^3\partial_{\rho} F_{x^0x^1}\big)\,+\,
{1\over \rho^2}\,\nabla^i\nabla_i\,F_{x^0x^1}\,+\,
R^2\,{f(\rho)\over \rho^3}\,{\cal P}^2
\,\varphi\,=\,0\,\,.
\label{eqF}
\eeq
This last equation, together with the equation of motion of $\varphi$ 
(the last equation in
(\ref{typeII})), constitute a system of coupled equations. In order 
to decouple them,
let us  define  the following combinations of $F_{x^0x^1}$ and $\varphi$:
\beq
\phi_{\pm}\equiv F_{x^0x^1}\pm {\cal P}\varphi\,\,.
\eeq
Notice that ${\cal P}=\sqrt{-\partial^2_{x^0}\,+\,\partial^2_{x^1}}$ 
makes sense acting on
a plane wave. It is straightforward to get the following system of decoupled
equations for $\phi_{\pm}$:
\beq
{R^4\over (\rho^2+L^2)^2}\,\partial^{\mu}\partial_{\mu}\,\phi_{\pm}\,+\,
{1\over \rho^3}\partial_{\rho}\big(\rho^3\partial_{\rho} \phi_{\pm}\big)\,+\,
{1\over \rho^2}\,\nabla^i\nabla_i\,\phi_{\pm}\,\pm\,
R^2\,{f(\rho)\over \rho^3}\,{\cal P}\phi_{\pm} \,=\,0\,\,.
\eeq
Let us  now expand $\phi_{\pm}$ in a basis of plane waves and 
spherical harmonics:
\beq
\phi_{\pm}\,=\,\xi_{\pm}(\rho)\,e^{ikx}\,Y^l(S^3)\,\,,
\eeq
where the product $kx$ in the plane wave is performed with the 
standard Minkowski metric.
Notice that
\beq
{\cal P}\,\phi_{\pm}\, =\,k_{01}\,\phi_{\pm}\,\,,
\eeq
where $k_{01}$ is defined as:
\beq
k_{01}\,\equiv \sqrt{(k_0)^2\,-\,(k_1)^2}\,\,.
\eeq
Moreover, in order to get rid of the factors $R$ and $L$ in the 
differential equations,  let us
define the following new variable:
\beq
\varrho\equiv {\rho\over L}\,\,,
\eeq
and the following rescaled quantities
\beq
\bar M^2\equiv-R^4\,L^{-2}\,k_{\mu}k^{\mu}\,\,,
\,\,\,\,\,\,\,\,\,\,\,\,
\bar k_{01}\equiv R^2\,L^{-1}\, k_{01}\,\,,
\,\,\,\,\,\,\,\,\,\,\,\,
\bar\Theta\,=\,\Theta\,L\,\,.
\eeq
Using these definitions it is an easy exercise to obtain the following
equation for the function $\xi_{\pm}$:
\beq
\partial_{\varrho}\big(\,\varrho^3\,\partial_{\varrho} \,\xi_{\pm}\,\big)\,+\,
\Bigg[\,\bar M^2\,{\varrho^3\over
(\varrho^2+1)^2}\,-\,l(l+2)\,\varrho\,\pm\,f(\varrho)\,\bar k_{01}\,
\Bigg]\,\xi_{\pm}\,=\,0\,\,,
\label{decoupled}
\eeq
where $f(\varrho)$ is given by:
\beq
f(\varrho)\,=\,\bar\Theta^2\,\partial_{\varrho}\,\Bigg[\,{2\varrho^2+1\over
1\,+\,\bar\Theta^4\,(\varrho^2\,+\,1)^2}\,\Bigg]\,\,.
\eeq
Notice that $\bar M$ and $\bar k_{01}$ are related as:
\beq
\bar k_{01}\,=\,\sqrt{\bar M^2\,+\,\bar k_{23}^2}\,\,,
\label{Mkrelation}
\eeq
where $\bar k_{23}$ is defined as
$\bar k_{23}\,\equiv\,R^2L^{-1}\,\sqrt{k_2^2+k_3^2}$. The explicit 
appearance of the momentum
$\bar k_{01}$ in the fluctuation equation (\ref{decoupled}) is a 
reflection of the breaking of
Lorentz invariance induced by the non-commutativity. We will show 
below that the admissible
solutions of eq. (\ref{decoupled}) only occur for some particular 
values of  $\bar M$ and $\bar
k_{01}$. Actually, for a given value of the momentum $\bar k_{23}$ in 
the non-commutative plane,
$\bar k_{01}$ depends on $\bar M$ (see eq. (\ref{Mkrelation})). Thus, 
the study of eq.
(\ref{decoupled}) will provide us with information about the mass 
spectrum of the theory and of
the dispersion relation satisfied by the corresponding modes.

When $\bar\Theta=0$, the equations (\ref{decoupled}) can be solved 
analytically in
terms of the hypergeometric function \cite{KMMW}. Indeed, let us 
define $\lambda$ as
$2\lambda+1=\sqrt{1+\bar M^2}$. Then, one can prove that the 
admissible solutions
have the form:
\beq
\xi_{\pm}(\rho)\,=\,\varrho^l\,\,(\,\varrho^2\,+\,1)^{-\lambda}\,\,
F(-\lambda, -\lambda+l+1;l+2;-\varrho^2)\,\,.
\label{commutative}
\eeq
The solutions (\ref{commutative}) behave at $\varrho\sim 0$ as
$\xi_{\pm}\sim \varrho^l$. Furthermore, the above solutions must 
vanish at infinity. This fact is
ensured by requiring that $-\lambda+l+1=-n$, with $n\in\ZZ_{+}$. This 
requirement implies the
following quantization condition  for $\bar M$\cite{KMMW}:
\beq
\bar M^2(\bar \Theta=0)\,=\,4\,(n+l+1)\,(n+l+2)\,\,.
\label{commutativeM}
\eeq
Due to the quantization condition imposed to $\lambda$, the power 
series of the hypergeometric
function terminates at order $\varrho^{2n}$ and, as a consequence, 
$\xi_{\pm}$ decreases as
$\varrho^{-(l+2)}$ as $\varrho\to\infty$.

\subsection{Study of the decoupled equations}
\begin{figure}
\centerline{\hskip -.8in \epsffile{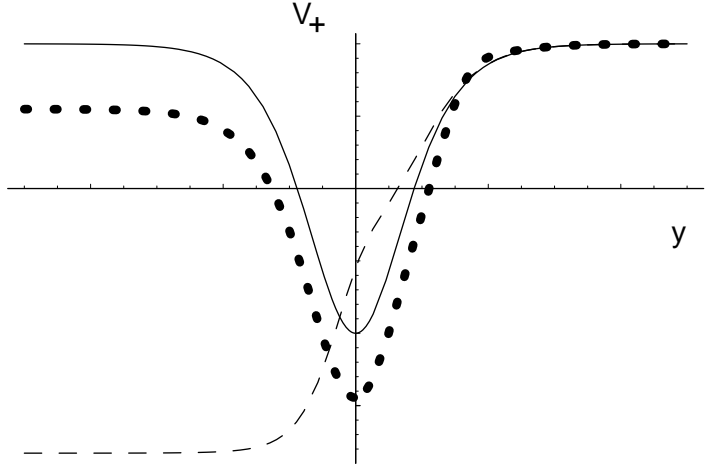}}
\caption{ The potential $V_+(y)$ for $\Theta=0$ (solid line) and for 
two non-vanishing values of
$\Theta$. The dashed line represents a potential such that there is 
only one turning point for
zero energy.}
\label{poten}
\end{figure}

We are now going to study the general features of the fluctuation 
equation (\ref{decoupled})
for a  general value of the non-commutativity parameter $\Theta$. In 
what follows the
fluctuation modes corresponding to $\xi_+$ and $\xi_-$ will be 
referred to simply as $+$ or
$-$modes respectively. A general technique to analyze equations such 
as those written in
(\ref{decoupled}) is by performing a  change of variables such  that
they can be written as the zero energy Schr\"odinger equation:
\beq
\partial_y^2\,\psi_{\pm}\,-\,V_{\pm} (y)\,\psi_{\pm}\,=\,0\,\,,
\label{Schrodinger}
\eeq
where $V_{\pm} (y)$ is some potential to be determined. The change of 
variables needed to pass
from  (\ref{decoupled}) to (\ref{Schrodinger}) is \cite{RS}:
\beq
e^y=\varrho\,\,,
\,\,\,\,\,\,\,\,\,\,\,\,\,\,\,\,\,\,\,\,\,\,\,\,\,\,\,
\psi_{\pm}\,=\,\varrho\,\xi_{\pm}\,\,.
\eeq
Notice that in this change of variables $\varrho\to\infty$ 
corresponds to $y\to+\infty$, while
the point $\varrho=0$ is mapped into $y=-\infty$. Furthermore,  by 
performing this change of
variables in eq. (\ref{decoupled}) one gets the following form of the potential
$V_{\pm} (y)$:
\bear
&&V_{\pm} (y)\,=\,-\bar M^2\,\,{e^{2y}\over 
(\,e^{2y}\,+\,1\,)^2}\,+\,(l+1)^2\,\mp\,
{4\bar\Theta^2 \bar k_{01}\over [1+\bar\Theta^4\,(\,e^{2y}\,+\,1\,)^2\,]^2}\,\,
[1\,-\,\bar\Theta^4e^{2y}(\,e^{2y}\,+\,1\,)\,]\,\,.\rc
\label{ncpotential}
\eear
Thus, the problem of finding the values of $\bar M$, as a function of 
$\bar k_{23}$, which give
rise to admissible solutions of eq. (\ref{decoupled}) can be 
rephrased as that of finding the
values of  $\bar M$ such that a zero-energy level for the potential 
(\ref{ncpotential}) exists.
When $\bar\Theta=0$, the potential (\ref{ncpotential}) represents a 
well centered around the
point $y=0$, where it is negative, while it becomes strictly positive 
at $y=\pm\infty$ (see
figure \ref{poten}).
  The Schr\"odinger equation in this commutative case has a discrete 
spectrum of bound states and
when  $\bar M$  is of the form  (\ref{commutativeM}) one of these 
bound states has zero energy.
When
$\Theta\not=0$ the shape of $V_{\pm}$ is deformed (see figure 
\ref{poten}) and it might happen
that there is no discrete bound state spectrum such that includes the 
zero-energy level. In
order to avoid this last possibility it is clear that one should have 
two turning points at
zero energy and, thus, the potential must be such that $\lim_{y\to 
\pm\infty}\,V_{\pm} (y)\ge
0$.  From the explicit expression of $V_{\pm} (y)$ given in eq. 
(\ref{ncpotential}) one readily
proves that:
\bear
&&\lim_{y\to +\infty}\,V_{\pm} (y)\,=\,(l+1)^2\,\,,\rc\rc
&&\lim_{y\to -\infty}\,V_{\pm} (y)\,=\,(l+1)^2\mp {4\bar\Theta^2 \bar 
k_{01}\over
(1+\bar\Theta^4)^2}\,\,.
\label{limits}
\eear
It is clear from (\ref{limits}) that the potential $V_{-}$ for the 
$-$modes is always positive
at $y=\pm\infty$. However, by inspecting the right-hand side of eq. 
(\ref{limits}) one easily
realizes this is not  the case for the
$+$modes. Actually,  from the condition $\lim_{y\to -\infty}\,V_{+} 
(y)\ge 0$ we get the
following upper bound on
$k_{01}$:
\beq
\bar k_{01}\,\le\,k_*(\bar\Theta)\,\,,
\quad\quad {\rm (+modes)}\,\,,
\label{bound}
\eeq
where the function $k_*(\bar\Theta)$ is defined as:
\beq
k_*(\bar\Theta)\,\equiv\,{(1+\bar\Theta^4)^2\over 
4\bar\Theta^2}\,\,(l+1)^2\,\,.
\label{k*}
\eeq
In  terms of the original unrescaled quantities, the above bound becomes:
\beq
k_{01}\,\le\,{L\over R^2}\,{(1+\Theta^4\,L^4)^2\over 4\Theta^2 
L^2}\,\,(l+1)^2\,\,.
\label{truebound}
\eeq

\begin{figure}
\centerline{\hskip -.8in \epsffile{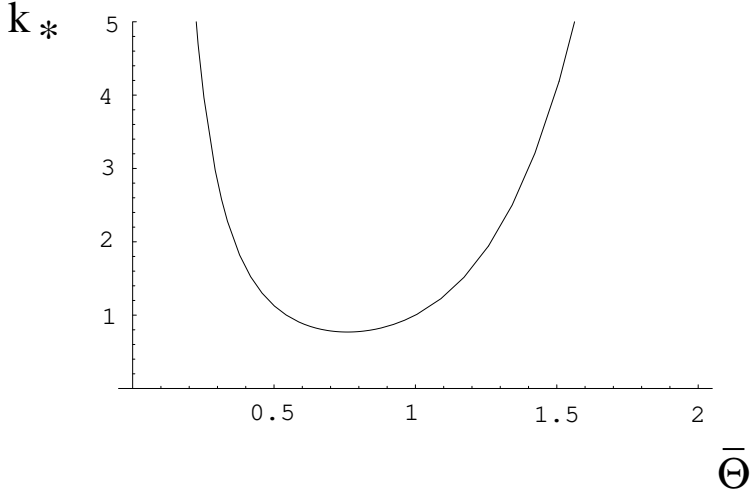}}
\caption{The function  $k_*(\bar\Theta)$  for $l=0$.}
\label{Theta}
\end{figure}

For a given value of the momentum $k_1$ along the $x^1$ direction, 
the above inequality
gives a bound on the energy $k_0$ of the $+$modes which resembles the 
stringy exclusion
principle. Indeed, as in string theory, we are dealing with a theory with a
fundamental length scale. In such theories one expects that it would be
imposible to explore distances smaller than the fundamental length which, in
turn, implies that some sort of upper bound on the energy and momentum must
hold. The function $k_*(\Theta)$ has been plotted in figure 
\ref{Theta}. Notice that this
upper bound is infinity when
$\Theta\to 0$ (as it should) but also grows infinitely when
$\Theta\to\infty$. Actually the function
$k_*(\bar\Theta)$ satisfies the following duality relation:
\beq
k_*(\bar\Theta)\,=\,\bar\Theta^4\,k_*({1\over \bar\Theta})\,\,.
\label{duality}
\eeq

A simple analysis of the function $k_*(\bar\Theta)$ reveals that it 
has a minimum
at a value of $\bar\Theta$ equal to
\beq
\bar\Theta_0\,=\,{1\over \root 4 \of 3}\,\approx\,0.7598\,\,,
\eeq
where it reaches the value
$k_*(\bar\Theta_0)\,=\,{4\over 3\sqrt{3}}\,(l+1)^2\approx 0.7698\,(l+1)^2$.

Notice that the upper bound (\ref{bound}) satisfied by the $+$modes 
implies, in particular, that
the spectrum of these modes is not an infinite tower as in the 
$\Theta=0$ case (see eq.
(\ref{commutativeM})). Instead, we will have a discrete set of values 
of $\bar M$ which, for a
given value of $l$, is parametrized by an integer $n$ such that $n\le 
n_*(\Theta)$, where
$n_*(\Theta)$ is a function of the non-commutative parameter which 
diverges when
$\Theta=0,\infty$. Actually, we will find below that, for some 
intermediate values of $\Theta$,
there are no $+$modes  satisfying the bound (\ref{bound}), \ie\ they 
disappear from the
spectrum.

Another interesting conclusion that one can extract from the analysis 
of the potential
$V_{\pm} (y)$ in eq. (\ref{ncpotential}) is the fact that for 
$\bar\Theta$ sufficiently large the
potential,  and therefore the spectrum, reduces to the one 
corresponding to the commutative
theory. Indeed, as shown in the plots of figure \ref{poten}, $V_{\pm} 
(y)$ is the same for all
values of $\bar\Theta$ in the region $y\to+\infty$. On the contrary, 
the limit of $V_{\pm}
(y)$ when $y\to-\infty$ does depend on $\bar\Theta$ (see eq. 
(\ref{limits})). However, it
follows from eq. (\ref{limits}) that when $\bar\Theta^6>>\bar k_{01}$ 
the $\bar\Theta$-dependent
term of $\lim_{y\to-\infty}\,V_{\pm} (y)$ can be neglected and, 
actually, when this condition
for $\bar\Theta$ is satisfied the form of
$V_{\pm} (y)$ in the whole range of $y$ is approximately the same as 
in the $\bar\Theta=0$ case.
We will verify below, both analytically and numerically,  that the spectrum for
$\bar\Theta\to\infty$ reduces to the one in the commutative theory. 
This fact, which might seem
surprising at first sight, is reminiscent of the Morita duality 
between irreducible modules
over the non-commutative torus (see \cite{NCFT} and references therein).

\setcounter{equation}{0}
\section{Meson spectrum}
In this section we will analyze in detail the solution of the 
differential equations
(\ref{decoupled}) for the $\pm$modes. The goal of this analysis is to 
determine the meson
spectrum of the corresponding non-commutative field theory at strong 
coupling. We will
first study this spectrum in the framework of the semiclassical WKB 
approximation, which has
been very successful \cite{MInahan} in the calculation of the 
glueball spectrum in the context of
the gauge/gravity correspondence \cite{glueball}. The WKB 
approximation is only reliable for
small $\Theta$ and large principal quantum number $n$, although in 
some cases it turns out to
give the exact result. In our case it would provide us of analytical 
expressions for the energy
levels, which will allow us to extract the main characteristics of 
the spectrum. We will confirm
and enlarge the WKB results by means of a numerical analysis of the 
differential equations
(\ref{decoupled}).

\subsection{WKB quantization}
The standard WKB quantization rule for the Schr\"odinger equation 
(\ref{Schrodinger}) is:
\beq
(n+{1\over 2})\pi\,=\,\int_{y_1}^{y_2}\,dy\,\sqrt{-V_{\pm}(y)}\,\,,
\,\,\,\,\,\,\,\,\,\,\,\,\,\,
n\ge 0\,\,,
\label{WKBquant}
\eeq
where $n\in\ZZ$ and  $y_1$ and $y_2$ are the turning points of the 
potential (\ref{ncpotential})
($V_{\pm}(y_1)=V_{\pm}(y_2)=0$).  One can evaluate the right-hand 
side of eq. (\ref{WKBquant})
by expanding it as a power series in $1/\bar M$. By keeping the 
leading and subleading terms of
this expansion, one can obtain $\bar M$ as a function of the quantum 
number $n$. The explicit
calculation for the potential (\ref{ncpotential}) (which makes use of 
the results of
\cite{MInahan, RS}) has been performed in appendix A. The expression 
of $\bar M$ one arrives at
is:
\beq
\bar M^2_{WKB}\,=\,4(n+1)\,\Bigg(n+l+1+(l+1)\sqrt{1\mp {\bar k_{01}\over
k_*(\bar\Theta)}}\,\,\Bigg)\,\,.
\label{WKBlevels}
\eeq
Notice that this spectrum only makes sense if the condition 
(\ref{bound}) is satisfied. When
$\bar\Theta=0$ the above formula is exact for $l=0$ and for 
non-vanishing $l$ it reproduces
exactly the quadratic and linear terms in $n$. In general, as we will 
check by comparing  it
with the numerical results,  eq. (\ref{WKBlevels})  is a good
approximation for small $\bar\Theta$ and  $n>> 2(l+1)$. Notice also 
that, due to the property
(\ref{duality}) of $k_*(\bar\Theta)$, the spectrum for 
$\bar\Theta\to\infty$ is identical to
that for
$\bar\Theta\to 0$.

Let us now  analyze some of the consequences of eq. (\ref{WKBlevels}).
Notice first of all that, as $\bar k_{01}$ and $\bar M$ are related as in eq.
(\ref{Mkrelation}), eq. (\ref{WKBlevels}) is really an equation which must be
solved to obtain $\bar M_{WKB}$ as a function of
$\bar k_{23}$ for given quantum numbers $n$ and $l$. Let us 
illustrate this fact when $n=l=\bar
k_{23}=0$. In this case $\bar k_{01}=\bar M$ and by a simple 
manipulation of eq.
(\ref{WKBlevels}) one can verify that
$\bar M_{WKB}$ is obtained by solving the equation:
\beq
F_{\pm}(\bar M_{WKB})\,\equiv\,{\bar M_{WKB}^3\over 16}\,-\,{\bar 
M_{WKB}\over 2}\,\pm\,
{1\over k_*(\bar\Theta)}\,=\,0,
\eeq
where the two signs correspond to those in eq. (\ref{WKBlevels}).
The function $F_{\pm}(\bar M_{WKB})$ has a unique minimum at a value of
$\bar M_{WKB}=\bar M_*=\sqrt{{8\over 3}}$, where it takes the value:
\beq
F_{\pm}(\bar M_*)\,=\,-{\sqrt{8}\over 3\sqrt{3}}\pm {1\over 
k_*(\bar\Theta)}\,\,.
\eeq
Obviously, the equation $F_{\pm}(\bar M_{WKB})=0$ has a solution for 
positive $\bar M_{WKB}$
iff  $F_{\pm}(\bar M_*)\le 0$. This condition is satisfied for all 
values of $\bar
\Theta$ for $F_-$, whereas for  $F_+$ the non-commutative parameter 
must be such
that
\beq
k_*(\bar\Theta)\,\ge {3\sqrt{3}\over \sqrt{8}}\,\,,
\quad\quad {\rm (+modes)}\,\,.
\eeq
 From the form of the function $k_*(\bar\Theta)$ we conclude that the above
inequality is satisfied when $\bar\Theta\le \bar\Theta_1$ and
$\bar\Theta\ge \bar\Theta_2$, where $\bar\Theta_1$ and $\bar\Theta_2$ 
are the two
solutions of the equation $k_*(\bar\Theta)=3\sqrt{3}/\sqrt{8}$. By numerical
calculation one obtains the following values of  $\bar\Theta_1$ and
$\bar\Theta_2$:
\beq
  \bar\Theta_1\,\approx\,0.376\,\,,
\,\,\,\,\,\,\,\,\,\,
\bar\Theta_2\,\approx\,1.239\,\,,
\,\,\,\,\,\,\,\,\,\,\,\,({\rm WKB})\,.
\label{WKBrange}
\eeq
Therefore, the ground state for the $+$modes disappears from the spectrum if
$\bar\Theta_1<\bar\Theta<\bar\Theta_2$. One can check similarly that  the
modes with $n>0$ also disappear if 
$\bar\Theta_1<\bar\Theta<\bar\Theta_2$ . Thus $(\bar\Theta_1,
\bar\Theta_2)$ is a forbidden interval of the non-commutative 
parameter for the $+$modes.
Moreover, notice that the condition (\ref{bound}) reduces in this 
$k_{23}=0$ case to the
inequality $\bar M\le k_*(\bar\Theta)$. One can check that when
$\bar\Theta\,=\bar\Theta_{1,2}$ the bound (\ref{bound}) is indeed satisfied,
since $\bar M=\sqrt{{8\over 3}}<k_*(\bar\Theta_{1,2})=3\sqrt{3}/\sqrt{8}$.

For a non-vanishing value of $\bar\Theta$ the energy levels of the 
$+$modes are cut off at some
maximal value $n_*(\bar \Theta)$ of the quantum number $n$. If we 
consider states with zero
momentum $\bar k_{23}$ in the non-commutative plane, for which $\bar 
M=\bar k_{01}$, the
function $n_*(\bar \Theta)$ can be obtained by solving the equation
\beq
  \bar M^2_{WKB}\Big|_{n=n_*(\bar \Theta)}\,=\,k_*(\bar\Theta)^2\,\,,
\label{cutoff}
\eeq
where the left-hand side is given by the solution of eq. 
(\ref{WKBlevels}). One can get an
estimate of $n_*(\bar \Theta)$ for small and  large values of $\bar 
\Theta$ by putting
$k_*(\bar\Theta)\to\infty$ on the left-hand side of eq. 
(\ref{cutoff}). In this case eq.
(\ref{WKBlevels}) yields immediately the value of $\bar M$ and if, 
moreover,  we
consider states with $l=0$, one arrives at the approximate equation:
\beq
4(\,n_*\,+\,1\,)\,(\,n_*\,+\,2\,)\,\approx\,k_*(\bar\Theta)^2\,\,,
\eeq
which can be easily solved, namely:
\beq
n_*(\bar \Theta)\approx -{3\over 2}\,+\,{1\over 
2}\,\sqrt{1\,+\,k_*(\bar\Theta)^2}\,\,.
\eeq

\subsection{Numerical results}
We would like now to explore numerically the spectrum of values of 
$\bar M$ for the differential
equation (\ref{decoupled}). With this purpose in mind, let us first
study the behavior of the solutions of the equation (\ref{decoupled}) for small
values of the radial variable $\varrho$. In particular we will try to find a
solution of the form:
\beq
\xi_{\pm}\sim \varrho^{\lambda}\,\,.
\label{xilambda}
\eeq
For small $\varrho$, the term containing $\bar M^2$ in eq. 
(\ref{decoupled}) can be
neglected and we find that $\varrho^{\lambda}$ is a solution of 
(\ref{decoupled}) if $\lambda$
satisfies the quadratic equation
\beq
\lambda^2\,+\,2\lambda\,-\,l(l+2)\,\pm\,{4\bar\Theta^2 \bar k_{01}\over
(1+\bar\Theta^4)^2}\,=\,0\,\,,
\label{lambdaeq}
\eeq
where the two signs correspond to those of eq. (\ref{decoupled}).
There are two solutions of the quadratic equation (\ref{lambdaeq}).
The solution  that corresponds in the $\bar\Theta=0$ limit to the function
(\ref{commutativeM}) is:
\beq
\lambda\,=\,-1+(l+1)\,\sqrt{1\mp {\bar k_{01}\over k_*(\bar\Theta)}}\,\,,
\label{lambdasol}
\eeq
where $k_*(\bar\Theta)$ is the function defined in eq. (\ref{k*}). 
Plugging the value of
these exponents  $\lambda$ on eq. (\ref{xilambda}), we obtain the 
behaviour of the
$\xi_{\pm}$ fluctuations in the IR, namely:
\beq
\xi_{\pm}\sim \varrho^{-1+(l+1)\,\sqrt{1\mp {\bar k_{01}\over 
k_*(\bar\Theta)}}}\,\,,
\qquad\qquad\qquad (\varrho\sim 0)\,\,.
\label{IRxi}
\eeq
As $k_*(\bar\Theta)\to \infty$ when  $\bar\Theta \to 0$,
it is straightforward to verify that eq. (\ref{IRxi}) reduces to 
$\xi_{\pm}\sim \varrho^l$
when  $\bar\Theta = 0$  and thus, as claimed above,  the IR behaviour 
of $\xi_{\pm}$
coincides with the one corresponding to the commutative fluctuations 
when the non-commutative
deformation is switched off. Moreover, it is interesting to notice 
that the condition  $\bar
k_{01}\le k_*(\bar\Theta)$ for the
$+$modes (eq. (\ref{bound})) appears naturally if we require 
$\xi_{+}$ to be real in the IR.

In
order to obtain the spectrum of values of $\bar M$, let us study the 
behaviour of $\xi_{\pm}$ for
large values of $\varrho$. When $\varrho\to \infty$, the terms 
containing $\bar M^2$ and
$\bar k_{01}$ in eq. (\ref{decoupled}) can be neglected and one can 
find a solution which
behaves as  $\xi_{\pm}\sim \varrho^{\Lambda}$ for large $\varrho$. An 
elementary calculation
shows that there are two possible solutions for the exponent 
$\Lambda$, namely $\Lambda=l,
-(l+2)$. Therefore, the general behavior of $\xi_{\pm}$ for large 
$\varrho$ will be of the form:
\beq
\xi_{\pm}\,\sim\,c_1\,\varrho^{l}\,+\,c_2\,\varrho^{-(l+2)}\,\,,
\qquad\qquad\qquad (\varrho\sim \infty)\,\,,
\label{UVxi}
\eeq
where $c_1$ and $c_2$ are constants. The allowed solutions are those 
which vanish at infinity,
\ie\ those for which the coefficient $c_1$ in (\ref{UVxi}) is zero. 
For a given value of the
momentum $\bar k_{23}$ in the non-commutative plane,
this condition only happens
for a discrete set of values of $\bar M$, which can be found 
numerically by solving the
differential equation (\ref{decoupled}) for a function which behaves 
as in eq. (\ref{IRxi}) near
$\varrho=0$ and then by applying the shooting technique to determine 
the values of $\bar M$ for
which $c_1=0$ in eq. (\ref{UVxi}). Proceeding in this way one gets a 
tower of values of $\bar
M$, which we will order according to the increasing value of $\bar 
M$. In agreement with the
general expectation for this type of boundary value problems, the fluctuation
$\xi_{\pm}$ corresponding to the $n^{th}$ mode has $n$ nodes. 
Moreover, as happened in the WKB
approximation, when $\bar \Theta\not =0,\infty$ the tower of states 
for the $+$modes terminates
at some maximal value $n=n_*$ and, for some values of $\bar\Theta$ 
and $\bar k_{23}$, there is
no solution for $\xi_{+}$ satisfying the boundary conditions and the 
bound (\ref{bound}) (see
below).

Let us now discuss the results of the numerical calculation when the 
momentum $\bar k_{23}$ in
the non-commutative direction is zero. In this case  one must put
$\bar k_{01}=\bar M$ in the differential equation (\ref{decoupled}) (see eq.
(\ref{Mkrelation})). For small $\bar \Theta$ the numerical results 
should be close to those
given by the WKB equation  (\ref{WKBlevels}). In order to check this 
fact we compare in the
table below the values  of $\bar M^2$ obtained numerically and those 
given by the WKB formula
(\ref{WKBlevels})  for the  $\pm$modes  for $l=0$ and $\bar \Theta=.1$

\bigskip\bigskip
\begin{tabular}[b]{|c|c|c|}
\hline
\multicolumn{3}{|c|}{
 \rule{0mm}{4.5mm}$\bar M^2$ for $+$modes for $l=0$, $\bar \Theta=.1$}\\
\hline
  $n$  &  Numerical & WKB \\
\hline
\ \ 0 & $7.37$  & $7.77$  \\
\ \ 1 & $22.24$  & $23.19$  \\
\ \ 2 & $44.67$  & $46.24$  \\
\ \ 3 & $74.76$  & $76.89$  \\
\ \ 4 & $112.52$  & $115.11$  \\
\ \ 5 & $157.94$  & $160.85$  \\
\hline
\end{tabular}
\qquad\qquad\qquad
\begin{tabular}[b]{|c|c|c|}
\hline
\multicolumn{3}{|c|}{\rule{0mm}{4.5mm}$\bar M^2$ for $-$modes for $l=0$,
$\bar \Theta=.1$}\\
\hline
  $n$  & Numerical   & WKB \\
\hline
\ \ 0 & $8.69$  & $8.22$  \\
\ \ 1 & $25.84$  & $24.76$  \\
\ \ 2 & $51.35$  & $49.58$  \\
\ \ 3 & $85.10$  & $82.68$  \\
\ \ 4 & $126.95$  & $124.05$  \\
\ \ 5 & $176.90$  & $173.66$  \\
\hline
\end{tabular}

\bigskip
We notice that, indeed,  the WKB values represent reasonably well the 
energy levels especially,
as it should, when the number $n$ is large.

\begin{figure}
\centerline{\hskip -.1in \epsffile{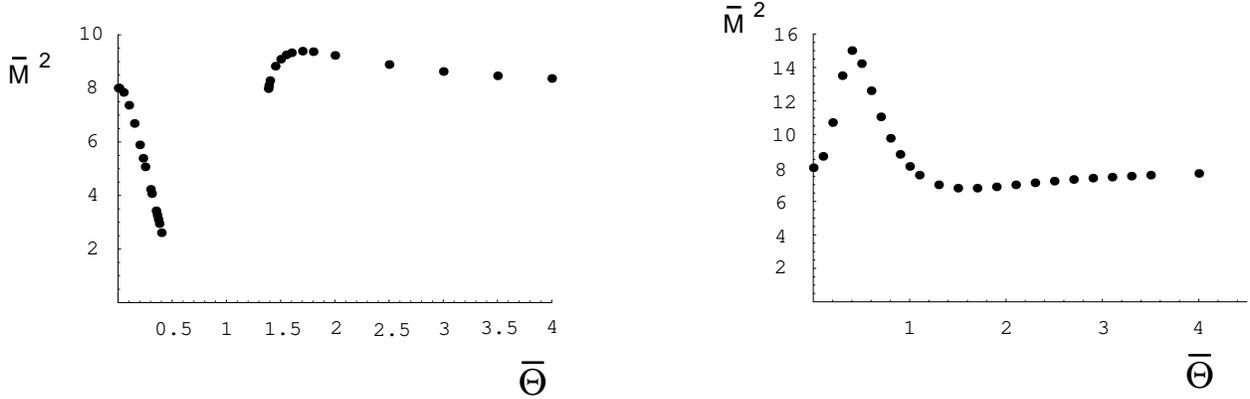}}
\caption{Numerical values of $\bar M^2$ for the ground state $n=l=0$ 
for $\bar k_{23}=0$ as a
function of $\bar \Theta$ for $\xi_+$ (left) and $\xi_-$(right). The 
corresponding value of the
commutative theory is $\bar M^2=8$.}
\label{spectrum}
\end{figure}

It is also interesting to analyze the behaviour of the ground state eigenvalue
$\bar M(n=0,l=0)$ with the non-commutative parameter $\bar\Theta$. In 
figure \ref{spectrum}
we plot our numerical results for the $\pm$modes when the momentum 
$\bar k_{23}$ is zero.
As expected, when $\bar\Theta=0$ we recover the value corresponding 
to the mass of the lightest
meson of the commutative theory. Moreover, if we increase 
$\bar\Theta$ the value of $\bar M$
for the $+$modes decreases, until we reach a  point from which there is no
solution of the boundary value problem satisfying the bound $\bar 
M\le k_*(\bar\Theta)$ in a
given interval  $\bar\Theta_1\le \bar\Theta\le \bar\Theta_2$ of the 
non-commutativity parameter.
The numerical values of $\bar\Theta_{1,2}$  are slightly different 
from the WKB result
(\ref{WKBrange}), namely $\bar\Theta_{1}=0.41$, 
$\bar\Theta_{2}=1.37$. It is interesting to notice
the jump and the different behaviour  of $\bar M$ at both sides of 
the forbidden region. Notice
also that, as expected, $\bar M$ approaches  the commutative result 
when $\bar \Theta$ is
very large.

For the $-$modes there is always a solution for the ground state for 
all values of $\bar\Theta$
and, again, the corresponding value of $\bar M$ equals the 
commutative result when
$\Theta=0,\infty$. Interestingly, the range of values of $\bar\Theta$ 
for which $\bar M$ differs
significantly from its commutative value is approximately the same as 
the forbidden interval for
the $+$modes (see figure \ref{spectrum}).

\begin{figure}
\centerline{\hskip -.1in \epsffile{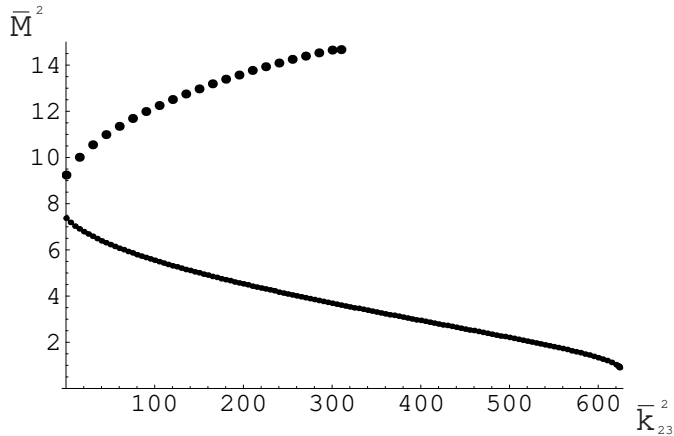}}
\caption{Numerical values of $\bar M^2$ for the ground state $n=l=0$ 
of the $+$modes  as a
function of $\bar k_{23}=0$ for two different values of $\bar 
\Theta$. The continuous line
corresponds to $\bar\Theta=0.1<\bar\Theta_1$, while the dotted line 
are the values of $\bar
M^2$ for $\bar\Theta=2> \bar\Theta_2 $.}
\label{spectrumk23}
\end{figure}

The differential equations for our fluctuations break explicitly the 
Lorentz invariance among
the Minkowski coordinates $x^0\cdots x^3$. In order to find out how 
this breaking is reflected
in the spectrum it is interesting to study the values of $\bar M^2$ 
for $\bar k_{23}\not=0$. In
this case $\bar k_{01}\not= \bar M$ (see eq. (\ref{Mkrelation})) and 
one can regard $\bar
k_{23}$ as a external parameter in our boundary value problem. When 
$\bar \Theta\not=0$ the
values of $\bar M$ that one obtains by solving the differential 
equation (\ref{decoupled}) do
depend on  $\bar k_{23}$. This dependence encodes the modification of 
the relativistic
dispersion relation due to the non-commutative deformation. For 
illustrative purposes let us
consider the spectrum for the $+$modes. The values of $\bar M^2$ as a 
function of $\bar k_{23}$
for the ground state ($n=l=0$) and two different values of 
$\bar\Theta$ are shown in figure
\ref{spectrumk23}. Notice that, as a consequence of  (\ref{bound}), 
the momentum $\bar k_{23}$ is
bounded from above. Moreover, $\bar M^2$ increases (decreases) with 
$\bar k_{23}$ when
$\bar\Theta < \bar\Theta_1$ ($\bar\Theta > \bar\Theta_2$), while it 
becomes independent of
$\bar k_{23}$ as $\Theta\to 0,\infty$.

\setcounter{equation}{0}
\section{Semiclassical strings in the non-commutative background}
\medskip

Let us now study the meson spectrum for large four-dimensional spin. 
An exact calculation of
this spectrum would require the analysis of the fluctuations of open 
strings attached to the
D7-brane in the non-trivial gravitational background of section 2. 
This calculation is not
feasible, even for the case of D7-branes in the $AdS_5\times S^5$ 
geometry considered in  ref.
\cite{KMMW}. However, for large four-dimensional spin we can treat 
the open strings
semiclassically, as suggested in ref. \cite{rotating}. In this 
approach one solves
the classical equations of motion of rotating open strings with 
appropriate boundary conditions.
These equations are rather complicated and can only be solved 
numerically. However, from these
numerical solutions we will be able to extract the energy and angular 
momentum of the string and
study how they are correlated. The analysis for the commutative 
$\Theta=0$ case was performed in
ref. \cite{KMMW}. Here we would like to explore the effect of a 
non-commutative deformation on
the behavior found in ref. \cite{KMMW}.

\subsection{Strings rotating in the non-commutative plane}
\label{counterclockwise}
As explained above, the spectrum of mesons with
large four-dimensional spin  can be obtained from classical
rotating open strings. In order to study these configurations
let us consider strings attached to the
D7 flavor brane, extended in the radial direction, and rotating in
the $x^2x^3$ plane (i.e. in the non-commutative directions). The
relevant part of the metric is:
\beq
{r^2\over
R^2}\left[-dt^2+h\left(\left(dx^2\right)^2+\left(dx^3\right)^2\right)\right]+{R^2\over
r^2}dr^2.
\eeq
Defining the coordinate $z$ as
\beq
z={R^2\over r}\,\,,
\eeq
and changing to polar
coordinates on the $x^2x^3$ plane, the  metric above becomes:
\beq
{R^2\over
z^2}\left[-dt^2+h\left(d\rho^2+\rho^2d\theta^2\right)+dz^2\right]\,\,,
\eeq
where $h$ and $B$, in terms of the new coordinate $z$, are
\bear
h&=&\left(1+R^8{\Theta^4\over z^4}\right)^{-1},\rc\rc
B&=&-\Theta^2{r^4\over R^2}h\,\rho\,d\rho\wedge d\theta=-\Theta^2{R^6\over
z^4}h\,\rho\,d\rho\wedge d\theta.
\eear
Notice that the coordinate $\rho$ introduced above as radial 
coordinate in the $x^2x^3$ plane
has nothing to do with the one defined in eq. (\ref{defrho}). 
Actually, since the string is
rotating around one of its middle points, we must allow $\rho$ to 
have negative values.
Let us now consider an open
fundamental string moving in the above background. Let 
$(\tau,\sigma)$ be worldsheet
coordinates. The embedding of the string will be determined by the 
functions $X^M(\sigma,\tau)$,
where $X^M$ represent the target space coordinates. Moreover, the 
dynamics of the string is
governed by the standard Nambu-Goto action
\beq
\label{rotact}
S=-{1\over 2\pi\alpha'}\int d\tau \,d\sigma \sqrt{-\det g}+{1\over
2\pi\alpha'}\int P[B]\,\,,
\eeq
where $g$ is the metric induced on the worldsheet of the string. Let 
us write the form of the
action (\ref{rotact}) for the following ansatz:
\beq
\label{rotans}
t=\tau \: ,\: \theta=\omega\tau \: ,\: \rho=\rho(\sigma) \: ,\:
z=z(\sigma),
\eeq
where $\omega$ is a constant angular velocity. If the prime denotes 
derivative with respect to
$\sigma$, the determinant of the induced metric and the pullback of 
the $B$ field for the ansatz
(\ref{rotans})  are given by:
\bear
\sqrt{-\det g}={R^2\over
z^2}\sqrt{\left(1-h\,\rho^2\omega^2\right)\left(h\,\rho'^2+z'^2\right)},\\
P[B]=\Theta^2{R^6\over z^4}h\,\rho\,\omega\,\rho'\,d\tau\wedge d\sigma\,\,.
\eear
By plugging this result on the  action (\ref{rotact}) one finds the 
following lagrangian
density:
\beq
\label{lgr}
{\cal L}={R^2\over
2\pi\alpha'}\left(-{1\over z^2}\sqrt{\left(1-h\,\rho^2\,\omega^2\right)
\left(h\,\rho'^2+z'^2\right)}+\Theta^2{R^4\over
z^4}h\,\rho\,\omega\,\rho'\right).
\eeq
The lagrangian (\ref{lgr}) does not depend explicitly on $t$ and 
$\theta$. Therefore, our system
has  two conserved quantities: the  energy $E$ and the (generalized) 
angular momentum $J$, whose
expressions are given by:
\bear
\label{Enrot}
E&=&\omega {\partial S \over \partial \omega} - S=
{R^2\over 2\pi\alpha'}\int d\sigma
{\sqrt{h \,\rho'^2+z'^2} \over z^2 \sqrt{1-h\,\rho^2\omega^2}} \, ,\rc
J&=&{\partial S \over \partial \omega}=
{R^2\over 2\pi\alpha'} \left[\int d\sigma \left({h\,\rho^2\omega \over
z^2} {\sqrt{h \,\rho'^2+z'^2} \over \sqrt{1-h\,\rho^2\omega^2}}+
\Theta^2{R^4\over
z^4}h\,\rho\,\rho'\right)\right]\,\,.
\eear
The physical angular momentum of the string is given by the first 
term on the second equation
in (\ref{Enrot}), namely:
\beq
J_1={R^2\over 2\pi\alpha'}\int
d\sigma
{h\rho^2\omega \over
z^2}{\sqrt{h\,\rho'^2+z'^2}\over\sqrt{1-h\,\rho^2\omega^2}}\,\,.
\label{J1}
\eeq

The equations of motion defining the time-independent profile of
the string can be obtained from (\ref{lgr}). Moreover, in addition one
must impose the boundary conditions that make the action stationary:
\beq
{\partial L\over \partial(X')^M}\delta X^M\Big|_{\partial\Sigma}=0\,\,.
\label{rotbc}
\eeq

As the endpoints of the string are
attached to the flavor brane placed at constant $z$, $\delta
z|_{\partial\Sigma}=0$. Moreover, since $\delta 
\rho\big|_{\partial\Sigma}$ is arbitrary, the
condition (\ref{rotbc}) reduces to:
\beq
{\partial {\cal  L}\over \partial \rho\,'}\Bigg|_{\partial\Sigma}=0\,\,.
\label{bcrho}
\eeq
Taking into account the explicit form of ${\cal  L}$ (eq. 
(\ref{lgr})), one can rewrite
eq. (\ref{bcrho}) as:
\beq
{\sqrt{1-h\,\rho^2\omega^2}\over \sqrt{h\,\rho'^2+z'^2}}\,\,\rho'
\,\Bigg|_{\partial\Sigma}={R^4\Theta^2\over 
z^2}\,\rho\omega\,\Bigg|_{\partial\Sigma}\,\,.
\label{bcrhodos}
\eeq
Eq. (\ref{bcrhodos}) can be used to find  the angle at which the 
string hits the flavor brane.
Indeed, let us suppose that the D7-brane is placed at $z=z_{D7}$ and 
that the string
intercepts the D7-brane at two points with coordinates 
$\rho=-\tilde\rho_{D7}<0$ and
$\rho=\rho_{D7}>0$. We will orient the string by considering the 
$\rho=-\tilde\rho_{D7}$
  ($\rho=\rho_{D7}$) end as its initial (final) point. It follows 
straightforwardly from eq.
(\ref{bcrhodos}) that the signs of $d\rho/d\sigma$ at the two ends of 
the string are:
\beq
{d\rho\over d\sigma}\,\Bigg|_{\rho=-\tilde\rho_{D7}}\,\le\,0\,\,,
\qquad\qquad\qquad
{d\rho\over d\sigma}\,\Bigg|_{\rho=\rho_{D7}}\,\ge\,0\,\,.
\label{signhit}
\eeq
Notice that for $\Theta=0$ the right-hand side
of eq. (\ref{bcrhodos}) vanishes, which means that 
$\rho'_{|\partial\Sigma}=0$ and, therefore,
the string ends orthogonally on the
D7-brane, in agreement with the results of ref. \cite{KMMW}. Moreover,
$\tilde\rho_{D7}\,=\,\rho_{D7}$ in the commutative case  and the 
string configuration is
symmetric around the point
$\rho=0$ (see ref. \cite{KMMW}). On the contrary,
eq. (\ref{bcrhodos}) shows that $\rho'_{|\partial\Sigma}$ does not vanish
in the non-commutative theory  and, thus,
the string hits the D7-brane at a certain angle, which depends on the 
non-commutative parameter
$\Theta$ and on the  $\rho$ and $z$ coordinates of $\partial\Sigma$. 
Actually, it follows from
the signs displayed in eq. (\ref{signhit}) that the string profile is 
not symmetric \footnote{Rotating strings which are not
symmetric with respect to a center of rotation were also considered
in \cite{talav} where the asymmetry came from considering
different quark masses.}
 around
$\rho=0$ when $\Theta\not= 0$ and that the string is tilted towards 
the region of negative
$\rho$\footnote{The strings rotating in the sense opposite to the one in
(\ref{rotans}), \ie\ with $\theta=-\omega\tau$, are tilted towards the region
of positive $\rho$. Apart from this, the other results in this section are not
modified if we change the sense of rotation.}. The actual values
of  $dz/d\rho$ at the two ends of the string can be obtained by 
solving the quadratic equation for
$z'/\rho'$ in (\ref{bcrhodos}). One gets:
\beq
\label{init}
{dz \over d\rho}\Bigg|_{\rho=-\tilde\rho_{D7}}=
-{\,z_{D7}^{2} \sqrt{1-\tilde\rho_{D7}^2 \,\omega^2}\over
\hat\Theta^2 \,\tilde\rho_{D7}\,\omega}\,\,,
\qquad\quad\qquad
{dz \over d\rho}\Bigg|_{\rho=\rho_{D7}}=
-{\,z_{D7}^{2} \sqrt{1-\rho_{D7}^2 \,\omega^2}\over
\hat\Theta^2 \,\rho_{D7}\,\omega}\,\,,
\eeq
where the sign of the right-hand side has been chosen to be in 
agreement with eq.
(\ref{signhit}) and we have defined $\hat \Theta=\Theta\,R^2$.

\begin{figure}
\centerline{\epsffile{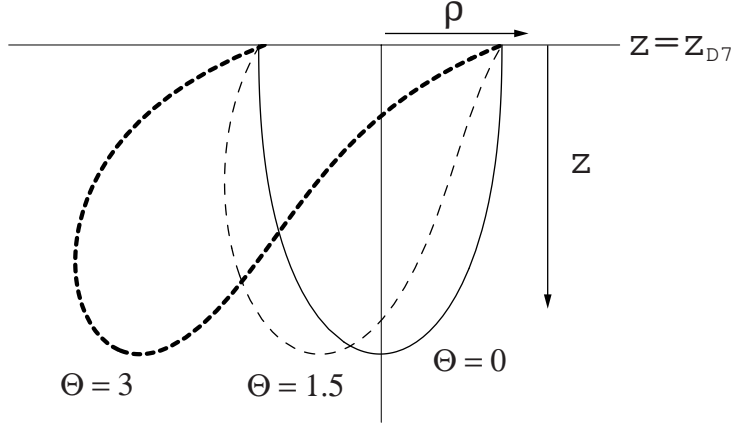}}
\caption{Rotating string profiles for three different values of the 
non-commutative parameter
($\Theta=$0, 1.5 and 3). The three curves correspond to the same 
value of $\rho_{D7}$. As
$\Theta$ grows the tilting increases. However, the maximal value of 
$z$ is roughly the same for
the three curves. }
\label{profile}
\end{figure}

Setting $\sigma=\rho$, the equation of motion defining the string
profile $z(\rho)$ can be easily obtained from the lagrangian (\ref{lgr}):
\bear
h{z''\over h+z'^2}-{h\,\rho\,\omega^2\over 1-h\,\rho^2 \omega^2}z'
-{4\hat  \Theta^2\over z^3}
h^2\rho\,\omega\sqrt{{h+z'^2\over 1-h\,\rho^2
\omega^2}}+\rc +{2h^2\over z\left(1-h\,\rho^2
\omega^2\right)\left(h+z'^2\right)}\Bigg\{z'^2 \left[1-{\hat
\Theta^4\over z^4} +\rho^2\omega^2\left(1-2h+{\hat  \Theta^4\over
z^4}h\right)\right]+
\rc+h\left[1-\rho^2\omega^2h\left(1-{\hat \Theta^4\over z^4}\right)\right]
\Bigg\} =0,
\label{rotdiffeq}
\eear
where now $z'={dz\over d\rho}$. In order to solve the second-order 
differential equation
(\ref{rotdiffeq}) we need to impose the value of $z$ and $z'$ at some 
value of $\rho$. Clearly,
the boundary condition (\ref{init}) fixes $z'$ at  $\rho=\rho_{D7}$. 
Since the string intersects
the D7-brane at this value of
$\rho$, it is evident that we have to impose also that:
\beq
z(\rho=\rho_{D7})=z_{D7}\,\,.
\label{zd7}
\eeq
By using the initial conditions (\ref{init}) at $\rho=\rho_{D7}$ and 
(\ref{zd7}), the equation of
motion (\ref{rotdiffeq}) can be numerically integrated for given 
values of $z_{D7}$, $\rho_{D7}$
and $\omega$. It turns out that these values are not uncorrelated. 
Indeed, we still have to
satisfy the condition written in eq. (\ref{init}) for
$z'(\rho=-\tilde\rho_{D7})$, which determines the angle at which the
$\rho=-\tilde\rho_{D7}$ end of the string hits the brane and it is 
not satisfied by arbitrary
values of $z_{D7}$, $\rho_{D7}$ and $\omega$. Actually, let us 
consider a fixed  value of
$z_{D7}$. Then,   by requiring  the fulfillment of eq. (\ref{init}), 
one gets a relation between
the quark-antiquark separation $\rho_{D7}+\tilde\rho_{D7}$ and the 
angular velocity $\omega$.
Before discussing this relation let us point out that, due to the 
tilting of the string, the
coordinate $\rho$ is not a good global  worldvolume coordinate in the 
region of negative  $\rho$,
since
$z(\rho)$ is a double-valued function in that region. In order to 
overcome this problem we will
solve eq. (\ref{rotdiffeq}) for $z(\rho)$ starting at 
$\rho=\rho_{D7}$ until a certain negative
value of $\rho$ and beyond that point we will continue the curve by 
parametrizing  the
string by means of a function
$\rho=\rho(z)$. The differential equation governing the function 
$\rho(z)$, which is similar to
the one written in (\ref{rotdiffeq}) for $z(\rho)$,   can be easily 
obtained from the lagrangian
density (\ref{lgr}) after taking $\sigma=z$. We have solved this 
equation by using as initial
conditions the values of the coordinate and slope of the last point 
of the $z=z(\rho)$ curve.
By performing the numerical integration in this way, the $\rho(z)$ 
curve is continued until
$z=z_{D7}$. Then $\tilde\rho_{D7}$ is determined as 
$-\tilde\rho_{D7}=\rho(z_{D7})$ and one can
check whether or not the string hits the flavor brane at 
$\rho=-\tilde\rho_{D7}$ with the angle
of eq. (\ref{init}). For a given value of the angular velocity 
$\omega$ this only happens for
some particular values of $\rho_{D7}$ and $\tilde\rho_{D7}$. Some of 
the profiles found by
numerical integration are shown in figure \ref{profile}. As explained 
above these curves are
tilted in general. This tilting increases with the non-commutative 
parameter and, for fixed
$\Theta\not=0$, it becomes more drastic as $\omega$ grows.

\begin{figure}
\centerline{\epsffile{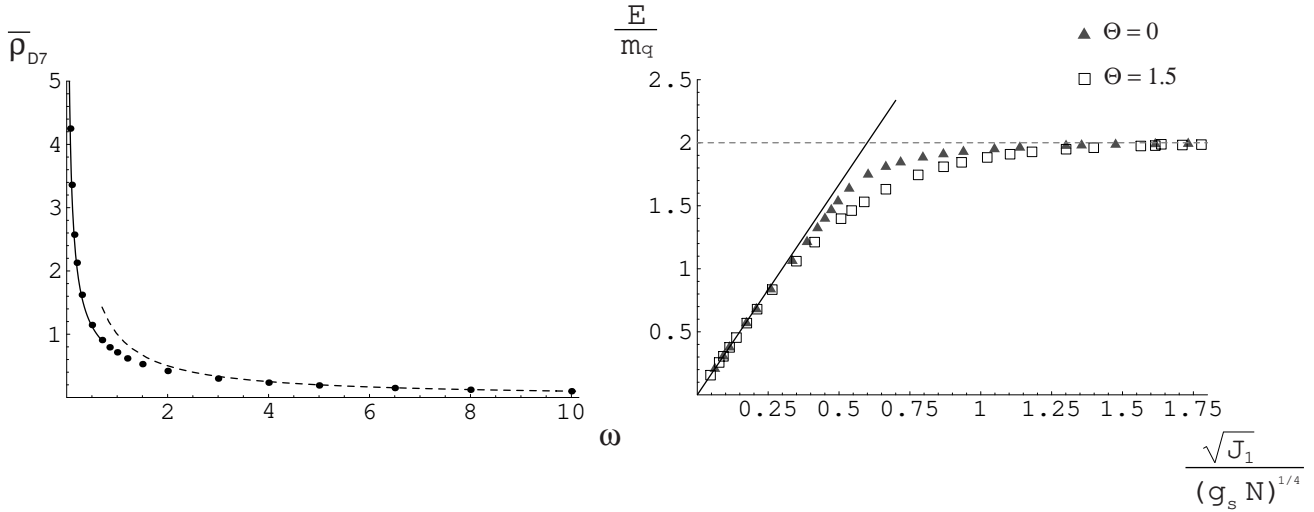}}
\caption{ On the left we plot the half of the quark-antiquark 
separation $\bar\rho_{D7}$ as a
function of the angular velocity $\omega$. For large $\omega$,
$\bar\rho_{D7}\approx \omega^{-1}$ (dashed line). In the $\omega\to 0$ region,
$\bar\rho_{D7}\sim \omega^{-2/3}$ (solid line). On the right plot we 
represent the
energy versus angular momentum of the rotating string for the 
commutative and non-commutative
theories. The meson masses follow Regge trajectories in the large $\omega$
(small $J_1$) region with an effective tension which does not depend on
the non-commutative parameter $\Theta$ and is given by eq. 
(\ref{tauR}). The linear Regge
trajectory given by eq. (\ref{linearRegge}) is also plotted for 
comparison (solid line).
For large $J_1$ (small
$\omega$) which corresponds to long strings, the energy approaches 
the free value $2m_q$.
We are setting $z_{D7}=R^2=\alpha'=1$.}
\label{numrot}
\end{figure}

As mentioned above, for a given value of $z_{D7}$ the fulfillment of 
the conditions written in
eq.  (\ref{init}) determines a relation between $\tilde\rho_{D7}$, 
$\rho_{D7}$ and
$\omega$. In order to characterize this relation let us define 
$\bar\rho_{D7}$ as the half of
the quark-antiquark separation, \ie\ 
$2\bar\rho_{D7}\,=\,\tilde\rho_{D7}+\rho_{D7}$.
In figure \ref{numrot} we have represented
$\bar\rho_{D7}$ as a function of $\omega$ for some non-vanishing 
value of $\Theta$. From these
numerical results one concludes that $\bar\rho_{D7}\to \infty$ when 
$\omega\to 0$, while
$\bar\rho_{D7}$ vanishes for large $\omega$. Actually, the behaviours 
found for small  and large
$\omega$ can be reproduced by a simple power law, namely:
\bear
&&\bar\rho_{D7}\,\sim \omega^{-{2\over 3}}\,\,,\quad\quad (\omega \to 
0)\,\,,\rc
&&\bar\rho_{D7}\,\sim \omega^{-{1}}\,\,,\quad\quad (\omega \to \infty)\,\,.
\label{powerrho}
\eear

The power law behaviours displayed in eq. (\ref{powerrho})  coincide 
with the ones found in ref.
\cite{KMMW} for the $AdS_5\times S^5$ background. They imply, in 
particular that $\omega\to 0$
corresponds to having long strings, whereas for $\omega\to\infty$ we 
are dealing with very short
strings.

Once
the profile of the string  is known, we can plug it on the right-hand 
side of eqs. (\ref{Enrot})
and (\ref{J1}) and obtain the energy and the angular
momentum  of the rotating string by numerical integration. As happens for the
$AdS_5\times S^5$ background, large(small) angular velocity 
corresponds to small (large) values
of the angular momentum $J_1$. Actually, for small $\omega$ the
angular momentum $J_1$ diverges, while $J_1\to 0$ for large $\omega$. 
The spectrum $E(J_1)$ can
be obtained parametrically by integrating the equation of motion for 
different values of
$\omega$. The results  for some values of $\Theta$ have been plotted 
in figure \ref{numrot}. For
small  $J_1$ (or large
$\omega$) the meson energy follows a Regge trajectory since the 
energy $E$ grows linearly with
$\sqrt{J_1}$. The actual value of the Regge slope can be obtained 
analytically (see the next
subsection). This Regge slope, which is nothing but an effective 
tension for the rotating
string, is independent of $\Theta$ (see figure \ref{numrot}). An 
explanation of this fact is
provided in the next subsection.  For intermediate values of $J_1$ 
the $E(J_1)$ curve does
depend on the non-commutative parameter while, on the
contrary, for large $J_1$ (or small $\omega$) one recovers again the 
$\Theta=0$ result, since the
energy becomes $2m_q$, with $m_q={R^2\over 2\pi\alpha'z_{D7}}$ being 
the mass of the quarks.
Actually this last result is quite natural since this large $J_1$ 
region corresponds to large
strings and one expects that the effects of the non-commutativity 
will disappear.

An interpretation of the behaviour of the spectrum in the two 
limiting regimes (small and large
$J_1$) was given in ref. \cite{KMMW}. Let us recall, and adapt to our 
system, the arguments of
\cite{KMMW}. For small $J_1$ the string is very short and it is not 
much influenced by the
background geometry. As  a consequence the spectrum is similar to the 
one in flat space, \ie\ it
follows a Regge trajectory with a tension which is just the proper 
tension $1/2\pi\alpha'$
appropriately red-shifted. This effective tension is independent of 
$\Theta$. However, we
will verify in section 6 that this is not the case when
computing the static quark-antiquark potential energy.
The static and dynamic  tensions are different and they only 
coincide for $\Theta=0$,
where we recover the results of ref. \cite{KMMW}.

\begin{figure}
\centerline{\epsffile{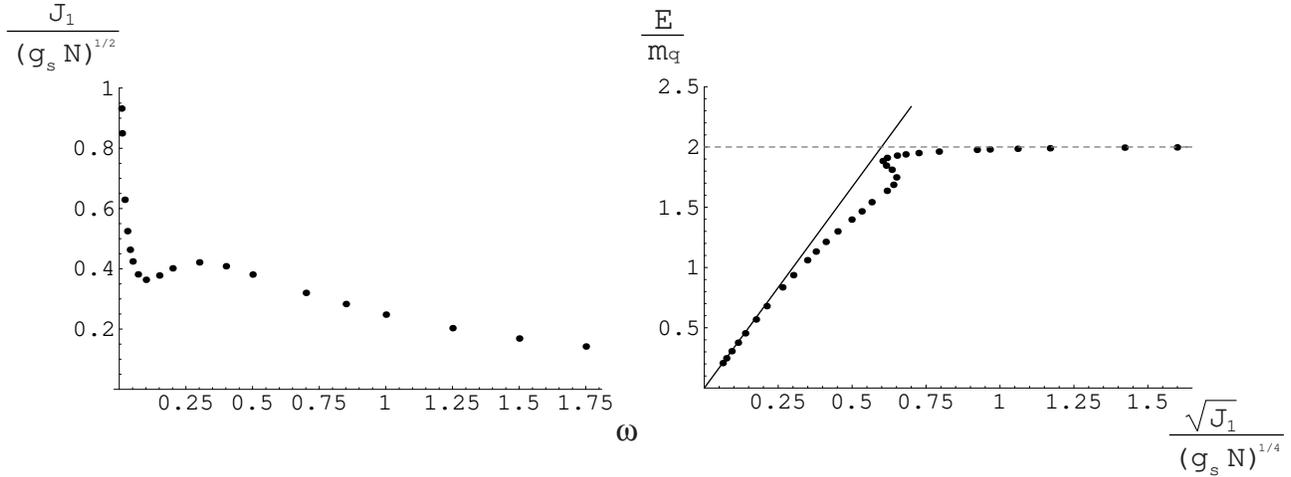}}
\caption{On the left we represent the angular momentum $J_1$ versus 
the angular velocity
$\omega$ for $\Theta=3$ and $z_{D7}=1$. The corresponding energy 
versus angular momentum plot is
shown on the right. The solid line corresponds to the linear Regge trajectory
(\ref{linearRegge}). }
\label{Largetheta}
\end{figure}

For large $J_1$ the spectrum corresponds to that of two non-relativistic
masses bound by a Coulomb potential. This is in agreement with the 
fact that in this long
distance regime the distance between the quark-antiquark pair is much 
larger than the inverse
mass of the lightest meson and one expects large screening 
corrections to the potential. We will
confirm this result by means of a static calculation, where we will 
compute the non-commutative
corrections to the large distance potential.

For $\Theta=0$ the $E(J_1)$ curve interpolates smoothly between the 
Regge behaviour at small
$J_1$ and the Coulomb regime at large $J_1$. However, when  $\Theta$ 
is non-vanishing  and
large enough,  the crossover region is  more involved since, in some 
interval of $J_1$, the
energy decreases with increasing angular momentum. By analyzing the 
numerical results of $J_1$
as a function of $\omega$ (see figure \ref{Largetheta}), one can 
easily conclude that this effect
is due to the fact that, when $\Theta$ is large, the angular momentum 
$J_1$ has some local
extremum for some intermediate values of
$\omega$.

\subsubsection{Large angular velocity}
\label{largeomega}

\begin{figure}
\centerline{\epsffile{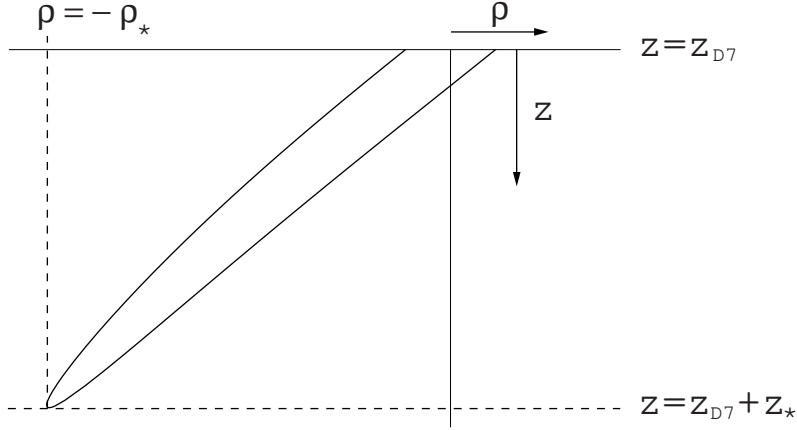}}
\caption{The string profile for large angular velocity. }
\label{parallel}
\end{figure}

Let us now analyze the case $\omega\to \infty$. As we have checked by numerical
computation, the solution in this
limit consists of a very short string with $\rho_{D7}=1/\omega$ and
$z(\rho)\approx z_{D7}$. Actually, for large $\omega$ the shape of 
the string resembles  two
nearly parallel straight lines joined at some turning point (see 
figure \ref{parallel}). Let
$\rho=-\rho_{*}$ be the $\rho$ coordinate of the turning point. The 
value of $\rho_{*}$ can be
approximately obtained by noticing that $d^2z/d\rho^2$ should diverge 
at $\rho=-\rho_{*}$. By
inspecting the differential equation (\ref{rotdiffeq}) one readily 
realizes that this can only
happen if  $1-h\rho^2\omega^2$ vanishes. Taking into account that the 
coordinate $z$ is very
close to $z_{D7}$ for these short strings, one concludes that:
\beq
\rho_{*}\,\approx\,{1\over h_{D7}^{{1\over 2}}\,\, \omega}\,\,,
\label{rho*}
\eeq
where
\beq
h_{D7}=\left(1+{\hat\Theta^4\over z_{D7}^4}\right)^{-1}\,\,.
\eeq
Let $z_{D7}+z_{*}$ be the value of the function $z(\rho)$ at 
$\rho=-\rho_*$ (see figure
\ref{parallel}). It is clear that $z_{*}$ measures how far the rotating 
string is extended in the
holographic direction. Our numerical calculations show that, for a 
given value of the angular
velocity $\omega$, the tilting of the string grows as $\Theta$ is 
increased but $z_{*}$ remains
nearly the same. Let us check this fact from the above equations for 
large values of
$\hat\Theta$. Indeed, in this case we can obtain approximately 
$z_{*}$ by multiplying the slope
(\ref{init}) by $\rho_{*}$, namely:
\beq
z_{*}\,\approx\,-{dz\over d\rho}\Bigg|_{\rho=\rho_{D7}}\rho_{*}\,\,.
\label{z*}
\eeq
Moreover, the expression (\ref{rho*}) of $\rho_{*}$ for large 
$\hat\Theta$ reduces to:
\beq
\rho_{*}\,\approx\,{\hat\Theta^2\over z_{D7}^2\omega}\,\,,
\label{rho*appr}
\eeq
and using (\ref{init}) to evaluate the right-hand side of eq. 
(\ref{z*}), one gets:
\beq
z_{*}\,\approx\,{\sqrt{1-\rho_{D7}^2\omega^2}\over \rho_{D7}\omega^2}\,\,.
\label{z*appr}
\eeq
Notice that all the $\hat\Theta$ dependence has dropped out from the 
right-hand side of eq.
(\ref{z*appr}) and, thus,  $z_{*}$ is independent of $\hat\Theta$ as 
claimed. Moreover, by
comparing with our numerical calculations we have found that eq. 
(\ref{z*appr}) represents
reasonably well $z_{*}$.

Our numerical results indicate that the energy $E$ and the angular 
momentum $J_1$ do not depend
on $\Theta$. Again we can check this fact by computing $E$ and $J_1$ 
for large $\Theta$. Notice
that, as $\rho_{D7}\approx 1/\omega$, one gets from (\ref{rho*appr}) that
$\rho_{D7}/\rho_{*}\approx z_{D7}^2/\hat\Theta^2$ and, therefore,
$\rho_{D7}<<\rho_{*}$ if $\hat\Theta$ is large and, as a consequence, 
the profile of the string
degenerates into two coinciding straight lines. Therefore,
the energy and the angular momentum in this regime can be obtained
by taking  $z=z_{D7}={\rm constant}$ in eqs. (\ref{Enrot}) and 
(\ref{J1}) and performing the
integration of $\rho$ between $\rho=0$ and $\rho =\rho_*$:
\beq
E\,\approx\,2\,{R^2\over 2\pi\alpha'}\,\,{h_{D7}^{{1\over 2}}\over 
z_{D7}^2}\,\,
\int_{0}^{\rho_*}\,\,{d\rho\over \sqrt{1-h_{D7}\,\rho^2\omega^2}}\,\,.
\eeq

By an elementary change of variables,
this integral can be done
analytically and, after taking into account the expression of 
$\rho_*$ (eq. (\ref{rho*})),  we
get the following expression of $E$:
\beq
E\approx {R^2\over 2\alpha'\omega\,z_{D7}^2}\,\,.
\label{ERegge}
\eeq
Notice that the dependence of $E$ on $h_{D7}$, and thus on $\Theta$, 
has disappeared.
Similarly, the angular momentum $J_1$ can be written as:
\beq
J_1\,\approx\,2\,{R^2\over 2\pi\alpha'}\,\,
{h_{D7}^{{3\over 2}}\,\omega\over z_{D7}^2}\,\,
\int_{0}^{\rho_*}\,\,d\rho{\rho^2\over \sqrt{1-h_{D7}\,\rho^2\omega^2}}\,=\,
{R^2\over 4\alpha'\omega^2\,z_{D7}^2}\,\,,
\label{JRegge}
\eeq
and $J_1$ is also independent of the non-commutative parameter.
Notice that $E$ and
$J_1$ depend on the angular velocity $\omega$ as $E\sim 1/\omega$ and 
$J_1\sim 1/\omega^2$. This
means that, indeed, $E\sim \sqrt{J_1}$ in this large $\omega$  regime 
and, actually, if we
define the effective tension for the rotating string as:
\beq
\tau_{eff}^R\,\equiv\,{E^2\over 2\pi J_1}\,\,,
\label{linearRegge}
\eeq
we have:
\bear
\tau_{eff}^R&=&{1\over 2\pi\alpha'_{eff}}\simeq
{R^2\over 2\pi\alpha'z_{D7}^2}\,\,.
\label{tauR}
\eear
As argued in ref. \cite{KMMW} for $\Theta=0$, the tension 
(\ref{tauR}) can be understood as the
proper tension ${1\over 2\pi\alpha'}$ at $z=z_{D7}$ which is then 
red-shifted  as
seen by a boundary observer.

\subsubsection{Small angular velocity}
 From the numerical computation displayed in figure \ref{numrot} one can
see that, in the large $J_1$ region (which corresponds to small $\omega$),
the energy becomes: $E\simeq 2m_q$, where $m_q={R^2\over
2\pi\alpha'z_{D7}}$ is the mass of the dynamical quarks. Furthermore, as
we can see in figure
\ref{numrot}, the separation between  the string endpoints $\bar\rho_{D7}$
when $\omega \to 0$ behaves as $\omega^{-2/3}$ (see eq. 
(\ref{powerrho})), which is the classical
result for two non-relativistic particles bound by a Coulomb 
potential, namely Kepler's law  \ie\
the cube of the radius is proportional to the square of the period.
This is precisely what one gets in the commutative case. In the next 
section we will calculate
the static quark-antiquark potential for long strings and we will 
verify that the dominant term
of this potential is of the Coulomb form, with a strength which is 
independent of $\Theta$.
Therefore, as expected on general grounds,
in this limit
the rotating string behaves exactly as in the commutative theory, and
this can be understood taking into account that this $\omega\to 0$ limit
corresponds to long strings, much larger than the non-commutativity scale
of the theory.

\setcounter{equation}{0}
\section{Hanging strings}
In this section we will evaluate the potential energy for a static 
quark-antiquark pair and we
will compare the result with the calculation of the energy for a 
rotating string presented in
section 5. Following the analysis of ref. \cite{Wilson}, let us 
consider a static configuration
consisting of a string stretched in the
$x^3$ direction with both ends attached to the $D7$-brane probe placed at
$r=r_{D7}$. The relevant part of the metric is:
\beq
{r^2\over R^2}\left[-dt^2+h\left(dx^3\right)^2\right]+{R^2\over
r^2}dr^2.
\eeq
Using $x^3$ as worldvolume coordinate and considering an ansatz of the form
$r=r(x^3)$, the Nambu-Goto action reads:
\beq
{\cal L}=-{1\over 2\pi\alpha'}\sqrt{-\det g}=-{1\over
2\pi\alpha'R^2}\sqrt{r^4h+R^4r'^2},
\eeq
where $r'$ denotes ${dr\over dx^3}$. As ${\cal L}$ does not depend 
explicitly on
$x^3$, the quantity
$r'{\partial {\cal L}
\over \partial r'} -{\cal L}$ is a constant of motion,\ie:
\beq
{r^4 h \over \sqrt{R^4r'^2 + r^4 h}}\,=\,{\rm constant}\,\,.
\label{conservation}
\eeq
\begin{figure}
\centerline{\epsffile{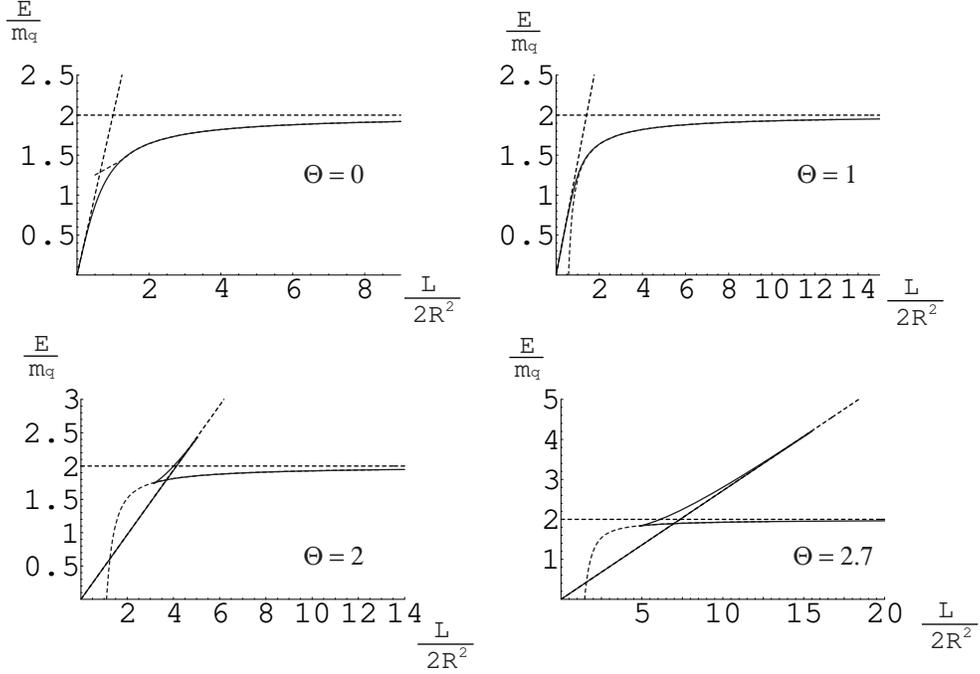}}
\caption{Energy of the static quark-antiquark configuration versus 
the string length $L$ for
$\Theta=0$ (commutative case), $\Theta=1$, $\Theta=2$, and 
$\Theta=2.7$ (setting
$z_{D7}=R^2=\alpha'=1$).
The solid line shows the result of the numeric computation, the straight
dashed line displays the Regge behaviour obtained in section
\ref{shortwilson}, and the curved dashed line shows the Coulomb potential
computed in section \ref{longwilson}.}
\label{numwilson}
\end{figure}
The configurations we are interested in are those in which the string 
is hanging of the flavor
brane at $r=r_{D7}$ and reaching a minimum value $r_0$ of the 
coordinate $r$. Since
$r'=0$ when $r=r_0$, we can immediately evaluate the constant of 
motion and rewrite eq.
(\ref{conservation}) as:
\beq
{r^4 h \over \sqrt{R^4r'^2 + r^4 h}}= {r_0^2 \over \sqrt{1+\Theta^4
r_0^4}}\,\, .
\eeq
 From this
expression we get readily $r'$ in terms of $r$, from which we can 
compute the string
length (i.e. the quark-antiquark separation) with the result:
\bear
L&=&2\int_{r_0}^{r_{D7}}{dx^3\over dr} dr = 2R^2 \int_{r_0}^{r_{D7}}
{r_0^2(1+\Theta^4 r^4) \over r^2\sqrt{r^4-r_0^4}} dr =\rc
&=&{2R^2 \over r_0}\int_1^{\frac{r_{D7}}{r_0}}{dy \over y^2 \sqrt{y^4-1}}+
2R^2r_0^3\,\Theta^4\,\int_1^{\frac{r_{D7}}{r_0}}{y^2 \over
\sqrt{y^4-1}}dy\,\,.
\label{length}
\eear
The energy for this static configuration becomes:
\beq
\label{energy}
E\,=\,-\int{\cal L}\,d\sigma\,=\,
{1\over \pi\alpha'}r_0\sqrt{1+\Theta^4
\,r_0^4}\int_1^{\frac{r_{D7}}{r_0}} {y^2 \over  \sqrt{y^4-1}}\,dy\,\,.
\eeq
These last two expressions  for $E$ and $L$ can be evaluated
numerically for any $r_0$ between $0$ and $r_{D7}$. From this result we obtain
the energy of the static quark-antiquark configuration as a
function of the distance between the quark and the antiquark. These results
have been plotted in figure \ref{numwilson}, where  one can observe that the
potential is linear for small quark-antiquark separation, while for 
large separation the energy
becomes constant and equal to $2m_q$. Notice that this behaviour is 
the same as the one we found
for the energy of the rotating strings. In the next two subsections 
we will analyze carefully the
small and long separation limits and we will obtain analytical 
expressions describing
the behaviour of the configuration in those limits.

\subsection{Short strings limit}
\label{shortwilson}
The small separation limit $L\to 0$ is achieved when $r_0\to
r_{D7}$. Then, by defining
${r_{D7}\over r_0}=1+\epsilon$ and making the change of variable
$y=1+z$, it is easy to get, at leading order:
\beq
L={2R^2 \over r_{D7}} \left(1+\Theta^4
r_{D7}^4\right)\epsilon^{\frac{1}{2}}\,\, ,
\qquad\qquad
E={r_{D7}\over \pi\alpha'}\sqrt{1+\Theta^4
r_{D7}^4}\,\epsilon^{\frac{1}{2}}\,\, .
\eeq
 From the dependence of $L$ and $E$  on $\epsilon$  we notice that, indeed,
$E=\tau_{eff}^W L$, where $\tau_{eff}^W$ is the effective tension 
 and is given by:
\beq
\tau_{eff}^W={E\over L}={r_{D7}^2 \over 2\pi\alpha'R^2}\,{1 \over 
\sqrt{1+\Theta^4
r_{D7}^4}}={R^2\over 2\pi\alpha'z_{D7}^2}\,{1 \over \sqrt{1+\Theta^4
r_{D7}^4}}\, .
\eeq
Notice that $\tau_{eff}^W$ decreases with the non-commutative 
parameter $\Theta$. Actually, the
maximal value of $\tau_{eff}^W$ is reached when $\Theta=0$, where it 
has the same value as the
tension $\tau_{eff}^R$  of the rapidly rotating string (see
eq. (\ref{tauR})). Therefore, for a general value of
$\Theta$ these two tensions have different values.

\subsection{Long strings limit}
\label{longwilson}
The limit $L\to\infty$ is achieved by taking the constant of
integration $r_0\to0$. It is easy to see from (\ref{length}),
(\ref{energy}) that the dependence on $\Theta$ cancels out
at leading order. Therefore, the Coulomb behaviour of the commutative 
theory found in ref.
\cite{KMMW} is exactly recovered at long distances. This is expected on
general grounds since, at distances bigger than the scale set by the
noncommutativity parameter of the field theory, the commutative theory
must be obtained. However, let us see what we get beyond the leading
term. Since $\frac{r_{D7}}{r_0}=\frac{2\pi\alpha' m_q}{r_0}$,  the 
length $L$ in (\ref{length})
can be rewritten as:
\beq
\label{longl}
L={2R^2 \over r_0}\left(\int_1^{\infty}{dy \over y^2
\sqrt{y^4-1}} -\int_{\frac{2\pi\alpha' m_q}{r_0}}^{\infty}\,{dy \over y^2
\sqrt{y^4-1}}\right)+
2R^2r_0^3\,\Theta^4\,\int_1^{\frac{2\pi\alpha' m_q}{r_0}}dy{y^2
\over
\sqrt{y^4-1}}\,.
\eeq
The first integral in eq. (\ref{longl}) can be explicitly computed, namely:
\beq
{\cal C}\equiv\int_1^{\infty}{dy \over y^2
\sqrt{y^4-1}}={\sqrt{\pi}\,\Gamma\left({3\over4}\right)\over
\Gamma\left({1\over4}\right)}\approx 0.59907\,\,.
\eeq
Moreover, when $\frac{r_{D7}}{r_0}=\frac{2\pi\alpha' m_q}{r_0}$  is 
large, the second integral
in (\ref{longl}) can be approximated as:
\beq
\int_{\frac{2\pi\alpha'
m_q}{r_0}}^{\infty}\,{dy
\over y^2
\sqrt{y^4-1}}\simeq{r_0^3\over 3\left(2\pi\alpha' m_q\right)^3}\,\,,
\eeq
while the third integral in (\ref{longl}) can be rewritten as:
\bear
\int_1^{\frac{2\pi\alpha' m_q}{r_0}}
dy{y^2\over\sqrt{y^4-1}}=
\frac{2\pi\alpha' m_q}{r_0}+\left[-1+\int_1^\infty dy
\left({y^2\over \sqrt{y^4-1}} -1\right)\right]-
\int_{\frac{2\pi\alpha' m_q}{r_0}}^\infty dy\left({y^2\over
\sqrt{y^4-1}} -1\right)\,\,.\rc
\label{thirdint}
\eear
The quantity in square brackets on the right-hand side of 
(\ref{thirdint}) is equal to
$-{\cal C}$, and approximating the integrand in the last term by
${1\over 2y^4}$, $L$ reads:
\beq
L={2R^2{\cal C}\over r_0}-{2R^2r_0^2\over 3\left(2\pi\alpha' m_q\right)^3}
+2\Theta^4R^2\left[2\pi\alpha'm_qr_0^2-{\cal C}r_0^3-{r_0^6\over
6\left(2\pi\alpha' m_q\right)^3} \right]\, .
\eeq
By solving iteratively for $r_0(L)$, one gets:
\beq
\label{r0}
r_0= {2R^2{\cal C}\over L}\left[1-\left({8R^6{\cal C}^2\over
3\left(2\pi\alpha' m_q\right)^3}-16\Theta^4R^6{\cal
C}^2\pi\alpha'm_q\right){1\over L^3}\right]+o\left(L^{-5}\right).
\eeq
On the other hand, for small $r_0$  the energy (\ref{energy}) becomes:
\bear
\label{enapp}
E&\simeq&{1\over \pi\alpha'}\left(r_0+{1\over 2}\Theta^4r_0^5\right)
\int_1^{\frac{2\pi\alpha' m_q}{r_0}}dy{y^2\over\sqrt{y^4-1}}=\rc
&=&{1\over \pi\alpha'}\left[2\pi\alpha'm_q-r_0{\cal C}-{r_0^4\over
6\left(2\pi\alpha' m_q\right)^3}+\Theta^4\pi\alpha'm_qr_0^4\right]+
o\left(r_0^5\right).
\eear
Substituting (\ref{r0}) into (\ref{enapp}) we get:
\beq
E= 2m_q+{2R^2{\cal C}^2\over \pi\alpha'}\left[-{1\over L}+{4R^6{\cal
C}^2\over
L^4}\left({1\over3\left(2\pi\alpha'
m_q\right)^3}-2\Theta^4\pi\alpha'm_q\right)\right]+o\left(L^{-5}\right).
\label{longenergy}
\eeq
The first term in (\ref{longenergy}) is just the rest mass of the 
quark-antiquark pair. The
$1/L$ term is the Coulomb energy whose strength, as anticipated 
above, does not depend on
$\Theta$ and is given by the expression obtained in ref. 
\cite{Wilson}. The last term in
(\ref{longenergy}) contains the $1/L^4$ corrections to the Coulomb 
energy, which depend on the
non-commutative parameter.

\subsection{Boosted hanging string}

As has already been pointed out, the non-commutativity parameter
explicitly breaks Lorentz invariance. Therefore, giving a velocity
to an object of the theory (and in particular to the string considered
in the previous subsections)
along the $x_2$ or $x_3$ cannot be described as a trivial Lorentz boost.
Thus, it is worth taking a brief glance to a moving hanging string
in the supergravity dual.

In fact, a moving string  can be coupled to the background
$B$-field, adding to the action a term similar to that of a charged
particle moving in a magnetic field.
Let us consider the following (consistent) ansatz for the
embedding of the worldsheet of the string in the target space:
\beq
t=\tau\,\,,\qquad
x_2=v\tau\,\,,\qquad
x_3=x_3(\sigma)\,\,,\qquad
r=r(\sigma)\,\,.
\eeq
By inserting this in the Nambu-Goto action (\ref{rotact}),
one readily obtains the effective lagrangian (we use the static
gauge $\sigma =x_3$):
\beq
{\cal L}=-\frac{1}{2\pi\alpha'R^2}
\left(\sqrt{1-h v^2}\sqrt{r^4h+R^4 r'^2}+\Theta^2 r^4 h v\right)\,\,.
\label{movingaction}
\eeq
It is interesting to notice the presence of the $h$ factor inside
the first square root. In usual commutative space-times, the fact that
the speed of light is the limiting velocity gets reflected in this
expression because the square root should be real. On the contrary, in
this case we have $v<h^{-\frac12}$ so the limit is larger than
one since $h < 1$. This agrees with the known fact that in non-commutative
theories it is possible to travel faster than light \cite{hashi}.
In fact, the usual causality condition changes and the light-cone
is substituted by a light-wedge \cite{gaume} (there is no causal limit
along the non-commutative directions). Notice that $h\to 0$ when
$\Theta \to \infty$ and also when $r\to \infty$. This last fact is
connected to the observation in \cite{chu}
where it was proved that the appearance of the light-wedge is related
to the UV of the field theory. A study of causality in the field
theory from the string dual has recently appeared in \cite{hubeny}.

Since (\ref{movingaction}) does not explicitly depend on $x_3$, one
can immediately obtain a first order condition that solves the
associated equations of motion $r'\frac{\partial {\cal L}}{\partial r'}
-{\cal L}=const$:
\beq
\sqrt{1-h v^2}\frac{r^4 h}{\sqrt{r^4 h+ R^4 r'^2}}
+ \Theta^2 r^4 h v = \sqrt{1-h_0 v^2} r_0^2 h_0^\frac12
+\Theta^2r_0^4h_0 v\,\,,
\eeq
where $r_0$ is the minimal value of $r$, where the string turns back up and
$h_0=h(r_0)$. It is straightforward to obtain the energy of this configuration:
\beq
E=\frac{1}{\pi\alpha'}\int_{r_0}^{r_{D7}} dr \frac{r^2 h^\frac12}{
\sqrt{r^4h(1-hv^2)-\left[r_0^2 h_0^\frac12\sqrt{1-h_0 v^2}
+\Theta^2v(r_0^4 h_0 -r^4 h)\right]^2}} \,\,.
\eeq
Of course, in the commutative limit $\Theta=0$, $h=h_0=1$, the
$v$-dependence factors out of the integral and one recovers the
usual relation $E=\frac{1}{\sqrt{1-v^2}} m_0$, where $m_0$ denotes the
energy for $v=0$. In the general non-commutative case, this relation
is modified in a complicated fashion since $v$ cannot be factored
out and the integral above cannot be solved analytically.

\section{Summary and conclusions}

In this paper we have studied, from several points of view, the 
addition of flavor degrees of
freedom to the supergravity dual of a gauge theory with
spatial non-conmutativity.  This analysis is carried out by
considering flavor branes in the probe approximation and by analyzing 
the dynamics of open strings attached to them. First, by using the Killing spinors of 
the gravity background and the kappa symmetry condition for the probes, we 
explicitly find the stable, supersymmetric embeddings for the flavor 
branes. They turn out to be the same as those of the conmutative theory
(this happens both for the background of ref. \cite{MR}  addressed in the main
text  and for the  non-commutative  dual of the Maldacena-Nu\~nez solution
studied in appendix B). Then,  by solving the equations for the 
fluctuations of the probe we have
computed the spectrum of scalar and vector mesons. We 
have found that the
effective metric ${\cal G}$  appearing in the quadratic lagrangian 
(\ref{quadraticBI})  which
governs these fluctuations is exactly the same as in the commutative 
theory. Recall that  ${\cal
G}$  is Lorentz invariant in the Minkowski directions and is the 
result of the combined action
of the background metric and
$B$ field on the Born-Infeld lagrangian of the probe. We have interpreted
${\cal G}$ as the open string metric relevant for our problem.

The Wess-Zumino part of the
action gives rise to the term (\ref{WZfluct}) which depends on 
$\Theta$, breaks explicitly the
Lorentz symmetry and, in addition, couples the scalar and vector 
fluctuations. This Wess-Zumino
term vanishes in the UV limit and, as a consequence, the UV dynamics 
of the fluctuations
does not depend on the non-commutative deformation. This is in sharp 
contrast with what happens
to the metric of the background, for which the introduction of the
non-commutative deformation  changes drastically the UV behaviour 
with respect to the
$AdS_5\times S^5$ geometry. By means of a change of variables we have 
been able to decouple the
differential equations of the fluctuations and we have studied the 
corresponding spectrum. By
looking at the $\Theta\to\infty$ limit of the equivalent 
Schr\"odinger problem we concluded that
the fluctuation spectrum for large $\Theta$ is the same as that 
corresponding to the $\Theta=0$
theory. We have also verified numerically that, for some intermediate 
values of $\Theta$, the
$+$modes are absent from the discrete spectrum.

We have also studied the configurations of open rotating strings with 
their two ends attached to
the D7-brane. The presence of a $B$ field changes the boundary 
conditions to be imposed at the
ends of the string and, as a consequence the string is tilted. Notice that
this tilting is also obtained when the strings are obtained as worldvolume
solitons of D-branes in the presence of the $B$ field (see, for example,
\cite{NCbion}).  By 
numerical integration we have
obtained the profile of the string and determined its energy 
spectrum. Short strings present a
Regge-like behaviour with an asymptotic slope which, despite of the 
tilting of the string, is
the same as in the commutative theory. On the other hand long strings 
behave as two
non-relativistic particles bound by a Coulomb potential, which is 
also the way in which they
behave in the commutative theory. Finally, in section 6, we 
have independently studied these two limits for a static string configuration
and briefly commented on the modification of the causality relation.

Let us finally mention some aspects that it would be worth to study
further. First of all it would be interesting to have a clearer understanding
of the fluctuation spectrum in the intermediate $\Theta$ region. Our results
seem to indicate that new physics could show up there. It is an open question
whether or not this is an artifact of our approach or  a real effect in the
field theory dual. By expanding the flavor brane action beyond second order
we would get a hint on the meson interactions, which we expect to depend on
the non-commutative parameter. Moreover, it would also be worth studying the
behaviour of glueballs in  non-commutative  backgrounds (see \cite{APT} for
some work along this line) in order to compare with the excitations coming
from the flavor sector. 
Naively, since glueballs are dual to 
closed strings and feel the collapsing of the metric in the UV, one would 
expect a very different behaviour from that found studying mesons, which, 
as explained above, are effectively embedded in the UV finite open string 
metric. Nevertheless, from the field theory point of view we do not expect
a priori such different behaviours. For this reason it  would be interesting
to clarify this point  further.

Another possible
extension of this work would be trying to incorporate dynamical baryons. It
was suggested in ref. \cite{KMMW} that such dynamical baryons could be
constructed from the baryon vertex. Moreover, according to the proposal of
ref. \cite{Ncbaryon}, the baryon vertex for the (D1,D3) background consists
of a D7-brane wrapped on a five-sphere and extended along the non-commutative
plane. Within this approach the dynamical baryons would be realized as
bundles of fundamental strings connecting the two types of D7-branes, namely
the flavor brane and the baryon vertex.

Non-commutative field theories have been formulated long time ago. They
display an intriguing UV/IR mixing whose implications are not fully
understood yet. The study of their non-perturbative structure might still
reserve us some surprises and we hope that our results could help to unveil
them.

\medskip
\section*{Acknowledgments}
\medskip
We are grateful to A. Cotrone, J. D. Edelstein,  D. Mateos, C. N\'u\~nez, J.
Russo, M. Schvellinger and K. Sfetsos
 for discussions and comments on the manuscript. The
work of D. A. and A. V. R. is supported in part by MCyT, FEDER and Xunta de
Galicia under grant  BFM2002-03881 and by  the EC Commission under  the FP5
grant HPRN-CT-2002-00325.
The work of A.~P. was
partially supported by INTAS grant, 03-51-6346, CNRS PICS $\#$
2530, RTN contracts MRTN-CT-2004-005104 and MRTN-CT-2004-503369
and by a European Union Excellence Grant, MEXT-CT-2003-509661.

\vskip 1cm
\renewcommand{\theequation}{\rm{A}.\arabic{equation}}
\setcounter{equation}{0}
\medskip
\appendix

\setcounter{equation}{0}
\section{WKB spectrum}
\medskip

In this appendix we will derive the WKB equation (\ref{WKBlevels}) 
for the mass spectrum. We
shall follow closely ref. \cite{RS} (see also ref. \cite{MInahan}). 
Let us suppose that
$\phi(\varrho)$ is a function which satisfies a differential equation 
of the form:
\beq
\partial_{\varrho}\,\big(\,g(\varrho)\,\partial_{\varrho}\,\phi\,\big)\,+\,
\big(\,\bar M^2\,q(\varrho)\,+\,p(\varrho)\,\big)\,\phi\,=\,0\,\,,
\label{ODE}
\eeq
where $\bar M$ is a mass parameter and $g(\varrho)$, $q(\varrho)$ and 
$p(\varrho)$ are three
arbitrary functions that are independent of $\bar M$. We will assume that
near $\varrho\approx 0,\infty$ these functions behave as:
\bear
&&g\approx g_1\varrho^{s_1}\,\,,
\,\,\,\,\,\,\,\,\,\,\,\,\,\,
q\approx q_1\varrho^{s_2}\,\,,
\,\,\,\,\,\,\,\,\,\,\,\,\,\,
p\approx p_1\varrho^{s_3}\,\,,
\,\,\,\,\,\,\,\,\,\,\,\,\,\,{\rm as}\,\,\varrho\to 0\,\,,\rc\rc
&&g\approx g_2\varrho^{r_1}\,\,,
\,\,\,\,\,\,\,\,\,\,\,\,\,\,
q\approx q_2\varrho^{r_2}\,\,,
\,\,\,\,\,\,\,\,\,\,\,\,\,\,
p\approx p_2\varrho^{r_3}\,\,,
\,\,\,\,\,\,\,\,\,\,\,\,\,\,{\rm as}\,\,\varrho\to \infty\,\,,
\eear
where $g_i$, $q_i$, $p_i$, $s_i$ and $r_i$ are constants. By a 
suitable change of variables the
general differential equation (\ref{ODE}) can be recast as a 
Schr\"odinger equation. The mass
spectrum can be computed by means of the WKB quantization rule 
(\ref{WKBquant}), where the
right-hand side is expanded in powers of $1/\bar M$. The values of 
$\bar M$ are determined from
the behaviour of $g$, $q$ and $p$ for $\varrho=0,\infty$. Let us 
determine this behaviour
in our case. The differential equations we are interested in have 
been written in eq.
(\ref{decoupled}). By comparing eqs. (\ref{decoupled}) and 
(\ref{ODE}) we immediately conclude
that $g$, $q$ and $p$  are given by:
\beq
g(\varrho)\,=\,\varrho^3\,\,,
\,\,\,\,\,\,\,\,\,\,\,\,\,\,
q(\varrho)\,=\,{\varrho^3\over (1+\varrho^2)^2}\,\,,
\,\,\,\,\,\,\,\,\,\,\,\,\,\,
p(\varrho)\,=\,-l(l+2)\varrho\,\pm\,\bar k_{01}\,f(\varrho)\,\,.
\label{gqp}
\eeq
By expanding  the functions written in eq. (\ref{gqp}) near 
$\varrho\approx 0$ we obtain:
\bear
&&g_1=1\,\,,
\,\,\,\,\,\,\,\,\,\,\,\,\,\,\,\,\,\,\,\,\,\,\,\,
\,\,\,\,\,\,\,\,\,\,\,\,\,\,\,\,\,\,\,\,\,\,\,\,
\,\,\,\,\,\,\,\,\,\,\,\,\,\,\,\,\,
s_1=3\,\,,\rc\rc
&&q_1=1\,\,,
\,\,\,\,\,\,\,\,\,\,\,\,\,\,\,\,\,\,\,\,\,\,\,\,
\,\,\,\,\,\,\,\,\,\,\,\,\,\,\,\,\,\,\,\,\,\,\,\,
\,\,\,\,\,\,\,\,\,\,\,\,\,\,\,\,\,
s_2=3\,\,,\rc\rc
&&p_1=-l(l+2)\,\pm\,{4\bar\Theta^2 \bar k_{01}\over (1+\bar\Theta^4)^2}
\,\,,
\,\,\,\,\,\,\,\,\,\,\,\,\,\,
s_3=1\,\,.
\label{IR}
\eear
Moreover, by studying the  functions (\ref{gqp}) at $\varrho\to\infty$ we
get:
\bear
&&g_2=1\,\,,
\,\,\,\,\,\,\,\,\,\,\,\,\,\,\,\,\,\,\,\,\,\,\,\,\,\,\,\,\,\,\,\,\,
r_1=3\,\,,\rc\rc
&&q_2=1\,\,,
\,\,\,\,\,\,\,\,\,\,\,\,\,\,\,\,\,\,\,\,\,\,\,\,\,\,\,\,\,\,\,\,\,
r_2=-1\,\,,\rc\rc
&&p_2=-l(l+2)\,\,,
\,\,\,\,\,\,\,\,\,\,\,\,\,\,
r_3=1\,\,.
\label{UV}
\eear

The consistency of the WKB approximation requires \cite{RS} that 
$s_2-s_1+2$ and $r_1-r_2-2$ be
strictly positive numbers, whereas $s_3-s_1+2$ and $r_1-r_3-2$ can be 
either positive or
zero. Notice that $s_2-s_1+2=r_1-r_2-2=2$ and $s_3-s_1+2=r_1-r_3-2=0$ 
and, thus, we are within
the range of applicability of the WKB approximation.  Let us define, 
following ref. \cite{RS},
the quantities
\beq
\alpha_1\,=\,s_2-s_1+2\,\,,
\,\,\,\,\,\,\,\,\,\,\,\,\,\,
\beta_1\,=\,r_1-r_2-2\,\,,
\eeq
and (as $s_3-s_1+2=r_1-r_3-2=0$ see \cite{RS}):
\beq
\alpha_2\,=\,\sqrt{(s_1-1)^2\,-\,4\,{p_1\over g_1}}\,\,,
\,\,\,\,\,\,\,\,\,\,\,\,\,\,\,\,\,\,\,\,\,\,\,\,\,\,\,\,
\beta_2\,=\,\sqrt{(r_1-1)^2\,-\,4\,{p_2\over g_2}}\,\,.
\eeq
The values of $\alpha_{1,2}$ and $\beta_{1,2}$ for our case can be 
straightforwardly obtained
from the results written in eqs. (\ref{IR}) and (\ref{UV}), namely:
\beq
\alpha_1\,=\,\beta_1\,=\,2\,\,,
\,\,\,\,\,\,\,\,\,\,\,\,\,\,
\alpha_2\,=\,2\,(l+1)\,\sqrt{1\mp {\bar k_{01}\over
k_*(\bar\Theta)}}\,\,,
\,\,\,\,\,\,\,\,\,\,\,\,\,\,
\beta_2\,=\,2(l+1)\,\,.
\label{alphabeta}
\eeq
The mass levels for large quantum number $n$ can be written in terms 
of $\alpha_{1,2}$ and
$\beta_{1,2}$ as \cite{RS}:
\beq
\bar M^2_{WKB}\,=\,{\pi^2\over 
\xi^2}\,(n+1)\,\bigg(n\,+\,{\alpha_2\over \alpha_1}\,+\,
{\beta_2\over \beta_1}\bigg)\,\,,\quad\quad (n\ge 0)\,\,,
\label{generallWKB}
\eeq
where $\xi$ is the following integral:
\beq
\xi\,=\,\int_0^{\infty} d\varrho\,\,\sqrt{{q(\varrho)\over g(\varrho)}}\,\,.
\eeq
By substituting the values of $q(\varrho)$ and $g(\varrho)$ for the 
case at hand (given in eq.
(\ref{gqp})) we obtain:
\beq
\xi\,=\,\int_0^{\infty} {d\varrho\over 1+\varrho^2}\,=\,{\pi\over 2}\,\,.
\eeq
Plugging this result in eq. (\ref{generallWKB}), and the values of 
$\alpha_{1,2}$ and
$\beta_{1,2}$ displayed in  eq. (\ref{alphabeta}), one readily gets
the WKB mass spectrum written in eq. (\ref{WKBlevels}).

\vskip 1cm
\renewcommand{\theequation}{\rm{B}.\arabic{equation}}
\setcounter{equation}{0}
\section{Flavor in the non-commutative Maldacena-Nu\~nez solution}
\medskip

In this section we are going to explore the possibility of adding 
flavor to the supergravity
dual of non-commutative ${\cal N}=1$ super Yang-Mills theory. The 
gravity dual of the
corresponding ${\cal N}=1$  commutative theory is the so-called 
Maldacena-Nu\~nez (MN) background
\cite{MN}, which is a geometry generated by a fivebrane wrapping a 
two-cycle. This geometry,
which was first obtained in \cite{CV} as a solution representing 
non-abelian magnetic monopoles
in four dimensions, is smooth and leads to confinement and chiral 
symmetry breaking (see ref.
\cite{MerReview} for a review). The mass spectrum of mesons in the MN 
background was obtained in
ref. \cite{flavoring} by considering the fluctuations of a D5-brane probe.

The non-commutative version of the MN background was obtained in ref.
\cite{NCMN} by means of a chain of strings dualities, very similar to 
the ones which led to
obtain the (D1,D3) bound state solution described in sect. 2. 
Actually, the background
found in \cite{NCMN} corresponds to a (D3,D5) bound state, with the 
D3-brane smeared in the
worldvolume of the D5, and wrapped on the two-cycle. Let us review in 
detail this solution.
The metric  in string frame is given by:
\beq
ds^2\,=\,e^{\phi}\,\,\Big[\,
dx^2_{0,1}\,+\,h^{-1}\,dx^2_{2,3}\,+\,
e^{2g}\,\big(\,d\theta_1^2+\sin^2\theta_1 d\phi_1^2\,\big)\,+\,
dr^2\,+\,{1\over 4}\,(w^i-A^i)^2\,\Big]\,\,,
\label{stringmetric}
\eeq
where $\phi$, $h$ and $g$ are functions of the radial coordinate $r$ 
(see below) and
$A^i$ is a one-form which can be written in terms of the angles 
$(\theta_1, \phi_1)$ and a
function $a(r)$ as follows:
\beq
A^1\,=\,-a(r) d\theta_1\,,
\,\,\,\,\,\,\,\,\,
A^2\,=\,a(r) \sin\theta_1 d\phi_1\,,
\,\,\,\,\,\,\,\,\,
A^3\,=\,- \cos\theta_1 d\phi_1\,.
\label{oneform}
\eeq
The  $w^i\,$'s appearing in eq. (\ref{stringmetric})
are  $su(2)$ left-invariant one-forms, satisfying
$dw^i\,=\,-{1\over 2}\,\epsilon_{ijk}\,w^j\wedge w^k$,
which can be represented in terms of three angles $\phi_2$, 
$\theta_2$ and $\psi$, namely:
\bear
w^1&=& \cos\psi d\theta_2\,+\,\sin\psi\sin\theta_2
d\varphi_2\,\,,\rc\rc
w^2&=&-\sin\psi d\theta_2\,+\,\cos\psi\sin\theta_2
d\phi_2\,\,,\rc\rc
w^3&=&d\psi\,+\,\cos\theta_2 d\phi_2\,\,.
\eear
The angles $\phi_i$, $\theta_i$ and $\psi$ take values in the range
$0\le \phi_i<2\pi$, $0\le \theta_i\leq\pi$ and $0\le \psi<4\pi$. Moreover,
the functions $a(r)$, $g(r)$ and  $\phi(r)$ are:
\bear
a(r)&=&{2r\over \sinh 2r}\,\,,\rc\rc
e^{2g}&=&r\coth 2r\,-\,{r^2\over \sinh^2 2r}\,-\,
{1\over 4}\,\,,\rc
e^{-2\phi}&=&e^{-2\phi_0}{2e^g\over \sinh 2r}\,\,,
\label{MNsol}
\eear
where $\phi_0$ is a constant ($\phi_0=\phi(r=0)$).
The function $h(r)$, which distinguishes in the metric the 
coordinates $x^2x^3$ from
$x^0x^1$, can be written in terms of the function $\phi(r)$  as follows:
\beq
h(r)\,=\,1\,+\,\Theta^2\,e^{2\phi}\,\,,
\label{MNh}
\eeq
where $\Theta$ is a constant which parametrizes the non-commutative 
deformation.

Let us denote by $\hat\phi$ the dilaton field of type IIB 
supergravity. For the solution
of ref. \cite{NCMN} this field takes the value:
\beq
e^{2\hat\phi}\,=\,e^{2\phi}\,h^{-1}\,\,.
\label{dilaton}
\eeq
Notice that, when the non-commutative parameter $\Theta$ is 
non-vanishing, the dilaton
$\hat\phi$ does not diverge at the UV boundary $r\to\infty$. Indeed, 
$e^{\hat\phi}$ reaches
its maximum value at infinity, where $e^{\hat\phi}\to \Theta^{-1}$. 
This behaviour is in sharp
contrast with the one corresponding to the commutative MN background, 
for which the dilaton
blows up at infinity.

This solution of the type IIB supergravity also includes a RR 
three-form $F_{(3)}$ given by:
\beq
F^{(3)}\,=\,-{1\over 4}\,\big(\,w^1-A^1\,\big)\wedge
\big(\,w^2-A^2\,\big)\wedge \big(\,w^3-A^3\,\big)\,+\,{1\over 4}\,\,
\sum_a\,F^a\wedge \big(\,w^a-A^a\,\big)\,\,,
\label{RRthreeform}
\eeq
where $F^a$ is the field strength of the su(2) gauge field $A^a$ of 
eq. (\ref{oneform}), defined
as:
\beq
F^a\,=\,dA^a\,+\,{1\over 2}\epsilon_{abc}\,A^b\wedge A^c\,\,.
\label{fieldstrenght}
\eeq
The different components of $F^a$ can be obtained by plugging the value of the
$A^a$'s on the right-hand side of eq. (\ref{fieldstrenght}). One gets:
\beq
F^1\,=\,-a'\,dr\wedge d\theta_1\,\,,
\,\,\,\,\,\,\,\,\,\,
F^2\,=\,a'\sin\theta_1 dr\wedge d\phi_1\,\,,
\,\,\,\,\,\,\,\,\,\,
F^3\,=\,(\,1-a^2\,)\,\sin\theta_1 d\theta_1\wedge d\phi_1\,\,,
\eeq
where the prime denotes derivative with respect to $r$. The NSNS $B$ field is
\beq
B\,=\,\Theta\,e^{2\phi}\,h^{-1}\,dx^2\wedge dx^3\,\,,
\label{NCMNB}
\eeq
and the corresponding three-form field strength $H=dB$ is
\beq
H=2\Theta\,\phi{\,'}\,e^{2\phi}\,h^{-2}\,dr\wedge dx^2\wedge dx^3\,\,.
\eeq
The solution has also a non-vanishing RR five-form $F^{(5)}$, whose 
expression is:
\beq
F^{(5)}\,=\,B\wedge F^{(3)}\,+\,{\rm Hodge \,\,\,dual}\,\,,
\label{NCMNF5}
\eeq
where $B$ and $F^{(3)}$ are given in eqs. (\ref{NCMNB}) and 
(\ref{RRthreeform}) respectively.
These RR field strengths satisfy the equations
\bear
&&dF^{(3)}\,=\,0\,\,,\rc\rc
&&dF^{(5)}\,=\,d{}^*F^{(5)}\,=\,H\wedge F^{(3)}\,\,,\rc\rc
&&d{}^*F^{(3)}\,=\,-H\wedge F^{(5)}\,\,.
\eear
Let us now define the seven-form $F^{(7)}$ as:
\beq
F^{(7)}\,=\,-{}^*F^{(3)}\,\,.
\eeq
Notice that, with this definition, all RR field strengths satisfy the equation
$dF^{(p)}\,=\,H\wedge F^{(p-3)}$. Then, they can be represented in 
terms of three
potentials $C^{(2)}$, $C^{(4)}$ and $C^{(6)}$ as follows:
\bear
&&F^{(3)}\,=\,dC^{(2)}\,\,,\rc\rc
&&F^{(5)}\,=\,dC^{(4)}\,-\,H\wedge C^{(2)}\,\,,\rc\rc
&&F^{(7)}\,=\,dC^{(6)}\,-\,H\wedge C^{(4)}\,\,.
\eear
We will need the values of these potentials in our calculations with 
flavor brane probes.
The expression of $C^{(2)}$ can be obtained from \cite{flavoring}, namely:
\bear
C^{(2)}&=&{1\over 4}\,\Big[\,\psi\,(\,\sin\theta_1 d\theta_1 \wedge 
d\phi_1 \,-\,
\sin\theta_2 d\theta_2\wedge d\phi_2\,)
\,-\,\cos\theta_1\cos\theta_2 d\phi_1\wedge d\phi_2\,-\rc\rc
&&-a\,(\,d\theta_1\wedge w^1\,-\,\sin\theta_1 d\phi_1\wedge w^2\,)\,\Big]\,\,.
\eear
In order to obtain the values of $C^{(4)}$ and $C^{(6)}$, let us 
introduce, following
\cite{flavoring}, the two-form ${\cal C}$, defined as:
\bear
{\cal C}&\equiv&-{e^{2\phi}\over 8}\,\,\Big[\,
\Big(\,(\,a^2-1\,)a^2\,e^{-2g}\,-\,16\,e^{2g}\,\Big)\,\cos\theta_1
d\phi_1\wedge dr
\,-\,(\,a^2-1\,)\,e^{-2g}\,w^3\wedge dr\,+\rc\rc
&&+\,a'\,\Big(\,\sin\theta_1 d\phi_1\wedge w^1\,+\,d\theta_1\wedge
w^2\,\Big)\,\Big]\,\,.
\label{calC}
\eear
Then, it can be checked that
\beq
{}^*\big(\,B\wedge F^{(3)}\,\big)\,=\,\Theta\,dx^0\wedge dx^1\wedge 
d{\cal C}\,\,,
\eeq
where the star denotes Hodge dual with respect to the metric 
(\ref{stringmetric}).
It follows from this result and the expression of $F^{(5)}$ given in 
eq. (\ref{NCMNF5})
that $C^{(4)}$ can be taken as:
\beq
C^{(4)}\,=\,B\wedge C^{(2)}\,+\,\Theta\,dx^0\wedge dx^1\wedge {\cal C}\,\,.
\label{MNC4}
\eeq
Similarly, one can verify that $F^{(7)}$ can be written as:
\beq
F^{(7)}\,=\,-h^{-1}\,dx^0\wedge\cdots\wedge dx^3\wedge d{\cal C}
\eeq
and, thus, it follows straightforwardly that the potential $C^{(6)}$ 
can be represented as:
\beq
C^{(6)}\,=\,-h^{-1}\,dx^0\wedge\cdots\wedge dx^3\wedge {\cal C}\,\,.
\label{C6}
\eeq
At this point it is worth to recall several interesting features of 
the non-commutative MN
background described above \cite{NCMN}. First of all, it is clear by 
inspecting the values of the
different fields that in the commutative limit $\Theta\to 0$ the 
function $h$ becomes one and we
smoothly recover the commutative MN solution. Moreover, in the deep 
IR limit $r\to 0$ the
background does not completely reduce to its commutative counterpart, 
since the RR five-form
$F^{(5)}$ is not zero for $r\to 0$. This fact is in contrast with the 
behaviour of the (D1,D3)
solution (see section 2) and has been interpreted in ref. \cite{NCMN} 
as an UV/IR mixing effect,
which is presumably absent in ${\cal N}=4$ super Yang-Mills but it is 
present in cases with less
supersymmetry.

\subsection{Killing spinors}

The flavor brane probes for the previous background are D5-branes 
extended along some calibrated
submanifold, which will be determined by using the kappa symmetry 
condition (\ref{kappa}). In
order to apply this technique one must previously know the Killing 
spinors of the background. In
this subsection we will determine these spinors for the solution of 
ref. \cite{NCMN}. Our
discussion will follow closely a similar analysis done in ref. 
\cite{flavoring} for the
commutative MN background. First of all, it is more convenient to 
work in Einstein frame, where
the metric  (\ref{stringmetric}) becomes:
\beq
ds^2_{E}\,=\,e^{{\phi\over 2}}\,\,h^{{1\over 4}}\,\Big[\,
dx^2_{0,1}\,+\,h^{-1}\,dx^2_{2,3}\,+\,
e^{2g}\,\big(\,d\theta_1^2+\sin^2\theta_1 d\phi_1^2\,\big)\,+\,
dr^2\,+\,{1\over 4}\,(w^i-A^i)^2\,\Big]\,\,.
\label{Einsteinmetric}
\eeq
We shall consider the following basis of frame one-forms:
\bear
&&e^{x^{0,1}}\,=\,e^{{\phi\over 4}}\,h^{{1\over 8}}\,dx^{0,1}\,\,,
\,\,\,\,\,\,\,\,\,\,\,\,\,\,\,\,\,\,\,\,\,\,\,\,\,\,
e^{x^{2,3}}\,=\,e^{{\phi\over 4}}\,h^{-{3\over 8}}\,dx^{2,3}\,\,,\rc\rc
&&e^{r}\,=\,e^{{\phi\over 4}}\,h^{{1\over 8}}\,dr\,\,,\rc\rc
&&e^1\,=\,e^{{\phi\over 4}}\,h^{{1\over 8}}\,e^g\,d\theta_1\,\,,
\,\,\,\,\,\,\,\,\,\,\,\,\,\,\,\,\,\,\,\,\,\,\,\,\,\,
e^2\,=\,e^{{\phi\over 4}}\,h^{{1\over 8}}\,e^g\,\sin\theta_1d\phi_1\,\,,\rc\rc
&&e^{\hat i}\,=\,{1\over 2}\,e^{{\phi\over 4}}\,h^{{1\over 8}}\,
(\,w^i\,-\,A^i\,)\,\,,
\,\,\,\,\,\,\,\,\,\,\,\,\,(i=1,2,3)\,\,.
\label{MNbasis}
\eear
Let us now define the following complex combination of
the NSNS and RR three-forms:
\beq
F\equiv e^{-{\hat \phi\over 2}}\,H\,+\,ie^{{\hat \phi\over 2}}\,F^{(3)}\,\,.
\eeq
In terms of the three-form $F$, the supersymmetric variation of the 
dilatino is \cite{SUSYIIB}:
\beq
\delta\lambda={i\over
2}\,\,\partial_{\mu}\,\hat\phi\,\Gamma^{\mu}\,\epsilon^*\,-\,
{i\over 24}\, F_{\mu_1\mu_2\mu_3}\,\Gamma^{\mu_1\mu_2\mu_3}\,\epsilon\,\,,
\label{deltalambda}
\eeq
while the gravitino variation is:
\beq
\delta\psi_{\mu}\,=\,D_{\mu}\,\epsilon\,+\,{i\over 1920}\,
F_{\mu_1\cdots\mu_5}^{(5)}\,\Gamma^{\mu_1\cdots\mu_5}\Gamma_{\mu}\epsilon\,+
\,{1\over 96} F_{\mu_1\mu_2\mu_3}\,
\big(\,\Gamma_{\mu}^{\,\,\,\mu_1\mu_2\mu_3}\,-\,
9\delta_{\mu}^{\mu_1}\,\,\Gamma^{\mu_2\mu_3}\,\big)\,\epsilon^{*}\,\,.
\label{deltapsi}
\eeq
The Killing spinors of the background are those for which the 
right-hand side of eqs.
(\ref{deltalambda}) and (\ref{deltapsi}) vanish. In order to satisfy 
the equations
$\delta\lambda=\delta\psi_{\mu}=0$ we will have to impose certain 
projection conditions on
$\epsilon$. First of all, we shall impose the same condition as in 
the commutative MN
background, namely \cite{flavoring}:
\beq
\Gamma_{12}\,\epsilon\,=\,\Gamma_{\hat 1 \hat 2}\,\epsilon\,\,,
\label{projectionone}
\eeq
where the $\Gamma$'s are flat Dirac matrices in the basis (\ref{MNbasis}).
We shall also introduce the  angle $\alpha$ which also appears  in 
the commutative case, namely
\beq
\cos\alpha\,=\,{\phi'\over 1+{e^{-2g}\over 4}\,(a^2-1)}\,\,,
\,\,\,\,\,\,\,\,\,\,\,
\sin\alpha\,=\,{1\over 2}\,
{e^{-g}a'\over 1+{e^{-2g}\over 4}\,(a^2-1)}\,\,,
\label{alpha}
\eeq
whose value can be obtained from the explicit form  (\ref{MNsol}) of 
the solution
\beq
\cos\alpha\,=\,{\rm \coth} 2r\,-\,{2r\over \sinh^22r}\,\,.
\label{alphaexplicit}
\eeq
Let us now define a new angle $\beta$ as:
\beq
\cos\beta\,=\,h^{-{1\over 2}}\,\,,
\,\,\,\,\,\,\,\,\,\,\,\,\,\,\,\,\,\,\,\,\,\,
\sin\beta\,=\,-\Theta\,e^{\phi}\,h^{-{1\over 2}}\,\,.
\label{MNbeta}
\eeq
Notice that $\beta=0$ when $\Theta=0$. Moreover,  from the definition 
of $h$ in (\ref{MNh}) one
can easily check that $\sin^2\beta+\cos^2\beta=1$. In terms of these angles
the condition $\delta\lambda=0$ takes the form:
\beq
\Gamma_{r\hat 1\hat 2\hat 3}\,\sigma_1\,\epsilon\,=\,
\Big[\,\cos\alpha\,(\cos\beta+\sin\beta\Gamma_{x^2x^3}\sigma_3)\,-\,
\sin\alpha\Gamma_{1\hat 1}\,\sigma_1\,\Big]\,\epsilon\,\,,
\label{projectiontwo}
\eeq
where we have written $\epsilon$ as a two-component real spinor.
In order to determine completely the Killing spinor, let us consider 
the variations of the
different components of the gravitino along the directions of the 
basis (\ref{MNbasis}).
First of all, the conditions $\delta\psi_{x^{\mu}}=0$ follow from the 
projections
(\ref{projectionone}) and (\ref{projectiontwo}). Moreover, 
$\delta\psi_{\hat i}=0$ is
satisfied if, in addition, the spinor $\epsilon$ satisfies
\beq
\sigma_1\epsilon\,=\,(\cos\beta\,+\,\sin\beta\Gamma_{x^2x^3}\sigma_3\,)\,\epsilon\,\,,
\label{sigmaprojection}
\eeq
which can be recast as
\beq
\sigma_1\epsilon\,=\,e^{\beta\,\Gamma_{x^2x^3}\sigma_3}\,\,\epsilon\,\,.
\eeq
Moreover, it can be checked that $\delta\psi_1=\delta\psi_2=0$ if we 
use the projections
(\ref{projectionone}), (\ref{projectiontwo}) and  (\ref{sigmaprojection})
and the first-order differential equations satisfied by  $g$ and $a$, namely:
\bear
g'&=&-{1\over 
2}\,(a^2-1)\,e^{-2g}\,\cos\alpha\,-\,ae^{-g}\,\sin\alpha\,\,,\rc\rc
a'&=&-2a\cos\alpha\,+\,e^{-g}\,(a^2-1)\,\sin\alpha\,\,.
\eear
Let us now solve the projections (\ref{projectionone}), 
(\ref{projectiontwo}) and
(\ref{sigmaprojection}). Notice,
first of all, that by using (\ref{sigmaprojection}) on the right-hand side of
(\ref{projectiontwo}), one arrives at:
\beq
\Gamma_{r\hat 1\hat 2\hat 3}\,\,\epsilon\,=\,
\Big(\,\cos\alpha\,-\, \sin\alpha\Gamma_{1\hat 1}\,\,\Big)\,\epsilon\,=\,
e^{-\alpha\Gamma_{1\hat 1}}\,\epsilon\,\,.
\label{projectionthree}
\eeq
Moreover,
since $[\,\Gamma_{x^2x^3}\sigma_3,\Gamma_{1\hat 1}\,]=
\{\sigma_1,\Gamma_{x^2x^3}\sigma_3\}=
\{\Gamma_{r\hat 1\hat 2\hat 3},\Gamma_{1\hat 1}\}=0$, we can solve 
(\ref{projectionone}),
(\ref{projectiontwo}) and (\ref{projectionthree}) as follows:
\beq
\epsilon\,=\,e^{{\alpha\over 2}\,\Gamma_{1\hat 1}}\,
e^{-{\beta\over 2}\,\Gamma_{x^2x^3}\sigma_3}\,\,\epsilon_0\,\,,
\eeq
where $\epsilon_0$ satisfies
\beq
\Gamma_{12}\,\epsilon_0\,=\,\hat\Gamma_{12}\,\epsilon_0\,\,,
\,\,\,\,\,\,\,\,\,\,\,\,
\sigma_1\epsilon_0\,=\,\epsilon_0\,\,,
\,\,\,\,\,\,\,\,\,\,\,\,
\Gamma_{r\hat 1\hat 2\hat 3}\,\,\epsilon_0\,=\,\epsilon_0\,\,.
\eeq
Plugging this expression into the equation $\delta\psi_r=0$, and 
using the fact that
the angles $\alpha$ and $\beta$ satisfy the following first-order equations
\beq
\alpha'\,=\,-e^{-g}\,a'\,\,,
\,\,\,\,\,\,\,\,\,\,\,\,\,\,\,\,\,\,\,\,\,\,\,\,
\beta'\,=\,-\Theta\,e^{\phi}\,\phi'\,h^{-1}\,\,,
\eeq
we get an equation which determines the radial dependence of $\epsilon_0$:
\beq
\partial_r\epsilon_0\,=\,{1\over 8}\,\phi'\,h^{-1}\,\epsilon_0\,+\,
{1\over 4}\,\Theta^2\,e^{2\phi}\,\phi'\,h^{-1}\,\epsilon_0\,\,.
\eeq
This equation can be easily integrated, namely:
\beq
\epsilon_0\,=\,e^{{\phi\over 8}}\,h^{{1\over 16}}\,\eta_0\,\,,
\eeq
with $\eta_0$ constant. Therefore, we can write the final form of $\epsilon$:
\bear
&&\epsilon\,=\,
e^{{\alpha\over 2}\,\Gamma_{1\hat 1}}\,
e^{-{\beta\over 2}\,\Gamma_{x^2x^3}\sigma_3}\,\,
e^{{\phi\over 8}}\,h^{{1\over 16}}\,\eta_0\,\,,\rc\rc
&&\Gamma_{12}\,\eta_0\,=\,\hat\Gamma_{12}\,\eta_0\,\,,
\,\,\,\,\,\,\,\,\,\,\,\,
\sigma_1\eta_0\,=\,\eta_0\,\,,
\,\,\,\,\,\,\,\,\,\,\,\,
\Gamma_{r\hat 1\hat 2\hat 3}\,\,\eta_0\,=\,\eta_0\,\,.
\label{NCMNspinor}
\eear
The algebraic conditions for $\eta_0$ written above determine four 
independent solutions, which is in agreement with the fact that our
background is dual to a  four-dimensional
${\cal N}=1$ theory. Notice that for $\Theta=0$ the above spinors 
coincide with the ones
obtained in ref. \cite{flavoring} for the commutative MN geometry. 
The effect on $\epsilon$ of
the non-commutative parameter is reflected in the extra rotation with 
angle $\beta$ on the
right-hand side of (\ref{NCMNspinor}) and in the additional power of 
$h$.  Moreover, being a
spinor of definite chirality of type IIB supergravity,
$\epsilon$ satisfies  $\Gamma_{x^0\cdots
x^3}\Gamma_{12}\Gamma_{r}\hat\Gamma_{123}\epsilon=\epsilon$. Using 
this fact in eq.
(\ref{projectionthree}), one can easily demonstrate that $\epsilon$ satisfies
\beq
\Gamma_{x^0\cdots x^3}\,\big(\,
\cos\alpha\,\Gamma_{12}\,+\,\sin\alpha\,\Gamma_1\hat\Gamma_2\,)\,
\epsilon\,=\,\epsilon\,\,,
\label{alphaproj}
\eeq
which will be useful in the next subsection.

\subsection{Supersymmetric D5-brane probes}
\medskip
In this subsection we want to characterize the embeddings of D5-brane 
probes which preserve the
same supersymmetry as the background. The form of these embeddings is 
determined by imposing the
kappa symmetry condition $\Gamma_{\kappa}\,\epsilon=\epsilon$, where 
$\epsilon$  is one of the
spinors (\ref{NCMNspinor}) and the general expression of 
$\Gamma_{\kappa}$ has been given in eq.
(\ref{generalgammak}). To parametrize the locus of the D5-brane
we shall use the following set of worldvolume coordinates
\beq
\xi^m\,=\,(x^0,\cdots,x^3,\theta_1,\phi_1)\,\,,
\eeq
and we will consider embeddings in which
\bear
&&\theta_2=\theta_2(\theta_1,\phi_1)\,\,,
\,\,\,\,\,\,\,\,\,\,\,\,\,\,\,\,\,\,\,\,\,\,\,\,
\phi_2=\phi_2(\theta_1,\phi_1)\,\,,\rc\rc
&&\psi=\psi(\theta_1,\phi_1)\,\,,
\,\,\,\,\,\,\,\,\,\,\,\,\,\,\,\,\,\,\,\,\,\,\,\,\,\,\,\,
r=r(\theta_1,\phi_1)\,\,.
\label{ansatz}
\eear

Moreover, we will take a vanishing  worldvolume gauge field $F=0$
and thus the value of the gauge-invariant combination ${\cal F}$, which
will be denoted by ${\cal F}^{(0)}$, will be  ${\cal 
F}^{(0)}\,=-\,P[B]$. From the expression of
$B$ in eq. (\ref{NCMNB}) it follows immediately  that  the only non-zero
component of $P[B]$ is:
\beq
{\cal F}_{x^2x^3}^{(0)}\,=\,-B_{x^2x^3}\,=\,-\Theta\,{e^{2\phi}\over h}\,\,.
\eeq
By adapting the general expression (\ref{generalgammak}) of 
$\Gamma_{\kappa}$ to this case, one
gets:
\beq
\Gamma_{\kappa}\,\epsilon\,=\,{1\over \sqrt{-\det(g_{ind}+{\cal 
F}^{(0)})}}\,\Gamma_{(0)}\,
[\,\sigma_1\,-\,\gamma^{x^2x^3}\,{\cal
F}_{x^2x^3}^{(0)}\,i\sigma_2\,]\epsilon\,\,,
\eeq
where $\Gamma_{(0)}$ has been defined in eq. (\ref{Gammazero}) and 
now we have denoted by
$g_{ind}$ to the induced metric on the worldvolume (which should not 
be confused with the
function $g$ appearing in the metric). For convenience we shall
work in the string frame\footnote{Notice that passing from the 
Einstein to the string frame
amounts to multiplying the spinor by some powers of $e^\phi$ and $h$. 
Thus, it follows that the
Killing spinors for the non-commutative MN background
in the string frame satisfy the same projections as those found in 
subsection B.1 in the
Einstein frame. },  where the metric is given by eq. (\ref{stringmetric}).
In this frame one has:
\beq
\gamma^{x^2x^3}\,=\,e^{-\phi}\,h\,\Gamma_{x^2x^3}\,\,,
\eeq
and, therefore, we can rewrite $\Gamma_{\kappa}\,\epsilon$ as
\beq
\Gamma_{\kappa}\,\epsilon\,=\,{1\over \sqrt{-\det(g_{ind}+{\cal 
F}^{(0)})}}\,\Gamma_{(0)}\,
[\,\sigma_1\,+\,\Theta\,e^{\phi}\Gamma_{x^2x^3}\,i\sigma_2\,]\epsilon\,\,.
\eeq
Since  $\tan \beta=-\Theta e^{\phi}$, where $\beta$ is the angle 
defined in  eq. (\ref{MNbeta}),
we can recast the above expression as:
\beq
\Gamma_{\kappa}\,\epsilon\,=\,{1\over \cos\beta 
\,\sqrt{-\det(g_{ind}+{\cal F}^{(0)})}}
\,\Gamma_{(0)}\,
[\,\cos\beta\,-\,\sin\beta\,\Gamma_{x^2x^3}\sigma_3\,]\,\sigma_1\,\epsilon\,\,.
\label{Gammakb}
\eeq
Moreover, by using eq. (\ref{sigmaprojection}) to evaluate the 
right-hand side of eq.
(\ref{Gammakb}), we get:
\beq
\Gamma_{\kappa}\,\epsilon\,=\,
{1\over \cos\beta \,\sqrt{-\det(g_{ind}+{\cal 
F}^{(0)})}}\,\Gamma_{(0)}\,\epsilon\,\,.
\label{Gammakc}
\eeq
Let us now evaluate $\det(g_{ind}+{\cal F}^{(0)})$. We will first 
compute the subdeterminant
corresponding to the $x^0\cdots x^3$ coordinates. In this 
subdeterminant the only off-diagonal
terms are those corresponding to ${\cal F}_{x^2x^3}^{(0)}$, and it is 
easy to prove that:
\beq
-\det(g_{ind}+{\cal F}^{(0)})_{x^0\cdots x^3}\,=\,e^{2\phi}\,
(\,e^{2\phi}\,h^{-2}\,+\,\Theta^2\,e^{4\phi}\,h^{-2}\,)\,=\,e^{4\phi}\,h^{-1}\,\,,
\eeq
where in the last step we have used the definition of $h$ in eq. (\ref{MNh}).
Thus, if $\hat g$ is the induced metric in the $(\theta_1, \phi_1)$ 
coordinates, one
gets:
\beq
\sqrt{-\det(g_{ind}+{\cal F}^{(0)})}\,=\,e^{2\phi}\,h^{-{1\over 
2}}\,\,\sqrt{\det{\hat g}}\,\,.
\label{determinant}
\eeq
Moreover, by using the string frame metric (\ref{stringmetric}), one 
concludes that:
\beq
\Gamma_{(0)}\,=\,e^{2\phi}\,h^{-1}\Gamma_{x^0\cdots 
x^3}\,\gamma_{\theta_1\phi_1}\,\,,
\label{G0}
\eeq
where $\gamma_{\theta_1\phi_1}$ is the antisymmetrized product of the 
two induced matrices
$\gamma_{\theta_1}$ and $\gamma_{\phi_1}$. Let us use eqs. 
(\ref{determinant}) and (\ref{G0}) to
evaluate the right-hand side of eq. (\ref{Gammakc}). After
taking into account that $\cos\beta=h^{-{1\over 2}}$, one gets:
\beq
\Gamma_{\kappa}\,\epsilon\,=\,{1\over \sqrt{\det{\hat g}}}\,\,
\Gamma_{x^0\cdots x^3}\,\gamma_{\theta_1\phi_1}\,\epsilon\,\,.
\label{Gammakd}
\eeq
Notice that the matrix acting on $\epsilon$ on the right-hand side of 
eq. (\ref{Gammakd}) does
not depend on $\Theta$. For an embedding of the form (\ref{ansatz}) 
one can compute
$\gamma_{\theta_1\phi_1}\,\epsilon$ by using the projections 
(\ref{projectionone}) and
(\ref{projectionthree}), which are also independent of $\Theta$. 
Then, the calculation of
$\gamma_{\theta_1\phi_1}\,\epsilon$ for the non-commutative MN 
background is exactly the same as
the one performed in ref. \cite{flavoring} for the $\Theta=0$ case. Moreover,
the kappa symmetry condition $\Gamma_{\kappa}\epsilon=\epsilon$ is 
satisfied by any Killing
spinor if it reduces to the equation (\ref{alphaproj}). This only 
happens if the embedding
functions (\ref{ansatz}) satisfy certain system of partial 
differential equations, which are
identical to the ones found and solved in ref. \cite{flavoring} for 
the $\Theta=0$ case. Thus,
despite the differences between the commutative and non-commutative 
MN backgrounds, the
supersymmetric D5-brane embeddings of the form (\ref{ansatz}) are 
exactly the same. As a further
check of this fact one can verify that
the equations of motion for configurations of the type (\ref{ansatz}) 
also reduce to the
ones of the $\Theta=0$ geometry. Notice that the lagrangian of the probe is
${\cal L}={\cal L}_{BI}\,+\,{\cal L}_{WZ}$, where the Born-Infeld and 
Wess-Zumino terms are
given by:
\bear
&&{\cal L}_{BI}\,=\,-e^{-\hat \phi}\,
\sqrt{-\det(g_{ind}+{\cal F})}\,\,,\rc\rc
&&{\cal L}_{WZ}\,=\,P[C^{(6)}]\,+\,P[C^{(4)}]\wedge {\cal F}\,+\,
{1\over 2}\,P[C^{(2)}]\wedge {\cal F}\wedge {\cal F}\,\,.
\label{MNlagrangian}
\eear
Let us denote by ${\cal L}^{(0)}={\cal L}_{BI}^{(0)}\,+\,{\cal 
L}_{WZ}^{(0)}$ the lagrangian for
a configuration of the type (\ref{ansatz})  with vanishing 
worldvolume gauge field.
If ${\cal F}={\cal F}^{(0)}=-P[B]$, we immediately prove, by using 
eqs. (\ref{determinant}) and
(\ref{dilaton}), that the BI term of ${\cal L}^{(0)}$ is :
\beq
{\cal L}_{BI}^{(0)}\,=\,-e^{\phi}\,\sqrt{\det \hat g}\,\,,
\eeq
which is the same as in the $\Theta=0$ MN background 
\cite{flavoring}. Let us consider next the
WZ term of ${\cal L}^{(0)}$. Since ${\cal F}^{(0)}\wedge {\cal 
F}^{(0)}=P[B]\wedge P[B]=0$,
one has:
\beq
{\cal L}_{WZ}^{(0)}\,=\,P[C^{(6)}]\,-\,P[B]\wedge P[C^{(4)}]\,\,.
\label{MNWZ0}
\eeq
Moreover, by using the expressions of $C^{(6)}$ and $C^{(4)}$  given in eqs. 
(\ref{C6}) and (\ref{MNC4}), one can prove that, if ${\cal C}$ is the two-form 
defined in (\ref{calC}), one
has:
\beq
{\cal L}_{WZ}^{(0)}\,=\,
-dx^0\wedge\cdots dx^3\wedge P[{\cal C}]\,\,,
\eeq
which has, indeed, the  same form as the WZ lagrangian for the 
commutative MN background in ref.
\cite{flavoring}. It follows that any embedding of the form 
(\ref{ansatz}) which solves the
equations of motion in the $\Theta=0$ case is also a solution of 
these equations in the
non-commutative background.

The solutions of the kappa symmetry condition 
$\Gamma_{\kappa}\epsilon=\epsilon$ have been
studied systematically in ref. \cite{flavoring}. Some of the 
embeddings found in ref.
\cite{flavoring} have all the right properties to be considered as 
flavor branes. They are
characterized by a function $r=r_0(\theta_1)$, which encodes the 
profile of the brane probe. The
explicit expression of the solution found in \cite{flavoring} is:
\bear
&&\theta_2=\theta_1\,\,,
\,\,\,\,\,\,\,\,\,\,\,\,\,\,\,
\phi_2=\phi_1\,\,,
\,\,\,\,\,\,\,\,\,\,\,\,\,\,\,
\psi=\pi,3\pi\,\,,\rc\rc
&&\sinh r_0\,=\,{\sinh r_*\over \sin\theta_1}\,\,,
\label{unperturbed}
\eear
where $r_*$ is a constant which represents the minimal value of the 
radial coordinate $r$.
There exists another physically equivalent embedding with 
$\theta_2=\pi-\theta_1$,
$\phi_2=2\pi-\phi_1$, $\psi=0,2\pi$ and with the same function 
$r_0(\theta_1)$ as in
(\ref{unperturbed}).

\subsection{Fluctuations}

Let us consider now fluctuations around the configuration 
(\ref{unperturbed}). Following
the approach of ref. \cite{flavoring}, we shall consider a configuration with
$\theta_2=\theta_1$, $\phi_2=\phi_1$ and $\psi=\pi,3\pi$, such that 
the radial coordinate
is given by
\beq
r(\theta_1,\phi_1,x)\,=\,r_0(\theta_1)\,+\,\chi(\theta_1,\phi_1,x)\,\,,
\label{rfluct}
\eeq
where $\chi$ is small and $r_0(\theta_1)$ is the function appearing in eq.
(\ref{unperturbed}). One can prove that the contribution of the BI 
term to the equations
of motion is the same as in the
$\Theta=0$ case.  In order to study the contribution of the WZ term, 
let us notice that,
for the angular embeddings we are considering, the pullback of the 
$w^i$ one-forms is:
\beq
P[w^1]\,=\,-d\theta_1\,\,,
\,\,\,\,\,\,\,\,\,\,\,\,\,\,\,
P[w^2]\,=\,-\sin\theta_1\,d\phi_1\,\,,
\,\,\,\,\,\,\,\,\,\,\,\,\,\,\,
P[w^3]\,=\,\cos\theta_1\,d\phi_1\,\,.
\eeq
Using this result, it is immediate to prove that
\beq
P[C^{(2)}]\,=\,0\,\,,
\eeq
and, therefore, the last term in ${\cal L}_{WZ}$  in 
(\ref{MNlagrangian}) vanishes. Using
this result in eq. (\ref{MNC4}), one gets that
\beq
P[C^{(4)}]\,=\,\Theta\, dx^0\wedge dx^1\wedge P[{\cal C}]\,\,,
\eeq
where
\beq
P[{\cal C}]\,=\,e^{2\phi}\,\Big[\,2 e^{2g}\,-\,{1\over 
8}\,(a^2-1)^2\,e^{-2g}\,\Big]\,
\cos\theta_1\,d\phi_1\wedge dr\,\,,
\eeq
with $dr$ being the exterior derivative of the function written in 
eq. (\ref{rfluct}). The
WZ term of the lagrangian can be written as
${\cal L}_{WZ}^{(0)}+\tilde{\cal L}_{WZ}$, where ${\cal 
L}_{WZ}^{(0)}$ is given in eq.
(\ref{MNWZ0}) and is the same that appears in the fluctuations of the 
$\Theta=0$ case.
The extra contribution $\tilde{\cal L}_{WZ}$ is given by
\beq
\tilde{\cal L}_{WZ}\,=\,P[C^{(4)}]\wedge F\,\,.
\eeq
By using the explicit form of $P[C^{(4)}]$, we can write $\tilde{\cal 
L}_{WZ}$ at second
order as:
\beq
\tilde{\cal L}_{WZ}\,=\,-\Theta\,f(\theta_1)\,\cos\theta_1\,\,\Big[\,
r_0'\,F_{x^2x^3}\,-\,\partial_{\theta_1}\chi\,F_{x^2x^3}\,-\,
\partial_{x^2}\chi\,F_{x^3\theta_1}\,-\,
\partial_{x^3}\chi\,F_{\theta_1x^2}\,\Big]\,\,,
\eeq
where
\beq
f(\theta_1)\equiv e^{2\phi}\,\big[\,2 e^{2g}\,-\,{1\over 8}\,
(a^2-1)^2\,e^{-2g}\,\big]\,_{\big|_{r=r_0(\theta_1)}}\,\,.
\eeq
The first term in $\tilde{\cal L}_{WZ}$ does not contribute to the 
equations of motion and
can be dropped. Moreover, by integrating by parts in the remaining 
terms and by using the
Bianchi identity
$\partial_{\theta_1}\,F_{x^2x^3}+\partial_{x^2}\,F_{x^3\theta_1}+
\partial_{x^3}\,F_{\theta_1x^2}=0$, one arrives at:
\beq
\tilde{\cal L}_{WZ}\,=\,\Theta\,\partial_{\theta_1}\big[\,f(\theta_1)\cos\theta_1\big]\,
\chi\,F_{x^2x^3}\,\,.
\eeq
This term in the lagrangian couples the scalar fluctuations $\chi$ to 
the gauge field
components $A_{x^2}$ and $A_{x^3}$, similarly to what happens in the (D1,D3)
background studied in the main text. However, in the present case the 
equations for the scalar
and vector fluctuations in the $\Theta=0$ case are very different and 
the procedure to decouple
them when $\Theta\not=0$ is far from obvious and will not be attempted here.

\end{document}